\documentclass[12pt]{iopart}
\usepackage{iopams,graphicx}
\begin{document}
\title{Transmission through a noisy network}
\author{Daniel Waltner$^{(1,3)}$ and  Uzy Smilansky$^{(1,2)}$}
\address{$^{(1)}$Department of Physics of Complex Systems, Weizmann Institute of Science, Rehovot 76100, Israel}
\address{$^{(2)}$School of Mathematics, Cardiff University, Cardiff, Wales, UK}
\address{$^{(3)}$Fakult\"at f\"ur Physik, Universit\"at Duisburg-Essen, 47048 Duisburg, Germany}
\begin{abstract}
Quantum graphs with leads to infinity serve as  convenient models for studying various aspects of systems which are usually attributed to chaotic scattering. 
They are also studied in several experimental systems and practical applications. In the present manuscript we investigate the effect of a time dependent random 
noise  on the transmission of such graphs, and in particular on the resonances which dominate the scattering observable such as e.g., the transmission and reflection 
intensities. We model the noise by  a potential $\alpha \delta (x-(x_0 +\gamma(t)))$ localized at an arbitrary point $x_0$ on any of the graph bonds, that fluctuates in 
time as a Brownian particle bounded in a harmonic potential described by the Ornstein-Uhlenbeck statistics. This statistics, which binds the Brownian motion within a 
finite interval, enables the use of a second order time-dependent perturbation theory, which can be applied whenever the strength parameter $\alpha$ is sufficiently 
small. The theoretical frame-work will 
be explained in full generality, and will be explicitly solved for a simple, yet nontrivial example.
\end{abstract}
\section{Introduction}
Quantum graphs provide very convenient models  for studying various phenomena which are usually associated with quantum chaotic systems \cite {KS}. In particular, they 
were proven to display universal spectral statistics \cite {Berkolaiko, AltGnutz} and transport properties \cite {KS-scat,Weidenmueller} which are characteristic of 
quantum chaos \cite {Bohigas,Haake}.  Quantum graphs are amply introduced and discussed in the literature, such as e.g., \cite {Gnutzmann} and \cite {BerkoKuch}.  
Here we shall be mainly concerned with their being a paradigm for quantum chaotic scattering, namely, the study of scattering through open quantum systems whose 
classical analogues display chaotic features \cite {BlumSmi}. Quantum graphs attract also experimental work, where the transmission through networks of quasi-one 
dimensional channels are considered. So far, most of the work was carried out with electro-magnetic microwaves \cite{Warsaw1,Warsaw2,Marbourg,Darmstadt}. However, 
networks of optical fibers or sound 
waves are also considered
\cite{Nir}.

Scattering on graphs occurs when some of the graph vertices are connected to infinite leads. Incoming waves on the leads are multiply reflected on the finite bonds in
the graph, until they emerge as outgoing waves which propagate to infinity on the leads. Stationary scattering theory can be applied  and an explicit expression for the 
unitary scattering matrix can be computed, based on the structure and  metric parameters of the graph and the vertex scattering matrices \cite{KS-scat}. The scattering 
within the finite part of the graph results in a complex sequence of resonances - typical for chaotic scattering of waves, but which can also display idiosyncratic effects such as the recently introduced "topological resonances" \cite{Schan,Waltner}.

In the present note we would like to extend the study of scattering through graphs, by examining the effect of noise modeled as a random, time-dependent perturbation 
coupled locally to the graph. This is motivated by experimental realizations of scattering graphs, which are affected by e.g., pickup of electromagnetic noise in 
microwave experiments. We expected that this  perturbation - even when week - is likely to affect the delicate balance of phases which are responsible for the emergence 
of narrow resonances.

The noise pick-up mechanism is modeled here by adding to the graph a delta potential on any of its edges,  at a position that fluctuates randomly in time. The position fluctuations are governed by an Ornstein-Uhlenbeck process \cite{Uhlenbbeck, Wang+Uhl,Risken}. That is, it performs a one-dimensional Brownian motion which is confined by a harmonic potential. Our treatment is perturbative in the strength of the added potential, we however do {\it{not}} assume the position of the potential to change slowly (adiabatically) in time.

The introduction of the time-dependent perturbation makes it necessary to address the problem using the time-dependent Schr\"odinger equation. One cannot use the 
scattering-matrix formalism, but rather compute the mean outgoing current which emerges at each of the outgoing edges when a constant flux of mono-energetic waves is scattered on the graph. As expected,
we find that the effect of noise is especially strong in the vicinity of the recently studied narrow topological resonances \cite{Schan,Waltner}, that arise from the weak coupling of quasi bound states on the graph to the outside.  We study  the effect of noise as a function of the noise memory time, and its coupling strength (to be chosen within the limitations set by our perturbation theory) and find quantitatively different behavior depending on these parameters. In this context we derive expressions for the current corrections that hold for an arbitrary open quantum graph and analyze them numerically for a special graph to illustrate the effect of the perturbation on resonances.

The outline of the article is as follows: In the next section we recall some methods common in  time dependent perturbation theory, and use them in a preparatory 
computation where the graph is taken as a single infinite line with a randomly fluctuating $\alpha\delta (x-\gamma(t))$ potential which induces the scattering.   
The general problem is discussed in section \ref{sect3},  where the fluctuating delta potential is positioned on one of the bonds of the graph. The essential features 
of the general theory are  studied in detail for a simple yet nontrivial graph - the loop graph - which displays narrow and broad resonances.  Section \ref{sect4} 
describes some numerical results where the effect of noise  on the width and position of the resonances is studied. In section \ref{sect5} we finally provide further 
details on the calculations leading to  the results given in sections \ref{sections4} and \ref{sect3}.

\section{Preliminaries: Scattering by a time dependent random potential on the line}\label{sect1}
\subsection{A Recursive solution of the time-dependent Schr\"odinger equation}
 The perturbative procedure used to solve the Schr\"odinger equation with a time dependent spatially and temporally localized scattering potential $V(x,t)$ is briefly 
 described in the following lines.  (For details see \cite{Alder}). We consider the Schr\"odinger equation
 \begin{equation}
\label{eq1}
i\hbar\frac{\partial}{\partial t}\Psi(x,t)=-\frac{\hbar^2}{2m}\frac{\partial^2}{\partial x^2}\Psi(x,t)+\alpha V(x,t)\Psi(x,t)
\end{equation}
with the parameter $\alpha$ that will be considered small and taken as an expansion parameter for our result. Our solution is subject to the boundary condition that at the starting time $t_0$  $\Psi(x,t_0)$ is obtained as solution of Eq.\ (\ref{eq1}) for $\alpha=0$.
We expand $\Psi(x,t)$ in terms of the stationary eigenfunctions $\Psi_k(x)$ obtained in the case $\alpha=0$
\begin{equation}
\label{eq2}
\Psi(x,t)=\int_{-\infty}^\infty d\omega dk{\rm e}^{-i\omega t}\beta_k(\omega)\Psi_k(x)
\end{equation}
with the expansion coefficients $\beta_k(\omega)$ to be determined perturbatively in $\alpha$. The integration variable $\omega$ contains here a small positive imaginary part that can be taken to zero at the end of the calculation in order to obtain only outgoing corrections to $\Psi(x,t)$. The stationary eigenfunctions $\Psi_k(x)$ considered above solve here the equation
\begin{equation}\label{Shro}
-\frac{\hbar^2}{2m}\frac{\partial^2}{\partial x^2}\Psi_k(x)=\frac{\hbar^2k^2}{2m}\Psi_k(x)
\end{equation}
with incoming boundary condition at $x\rightarrow\infty$.
Inserting the expansion (\ref{eq2}) into (\ref{eq1}), we obtain with (\ref{Shro})
\begin{eqnarray}
\int_{-\infty}^\infty d\omega dk\, \hbar\omega{\rm e}^{-i\omega t}\beta_k(\omega)\Psi_k(x)&&=\int_{-\infty}^\infty d\omega dk \frac{\hbar^2k^2}{2m}{\rm e}^{-i\omega t}\beta_k(\omega)\Psi_k(x) \nonumber\\&&+\alpha\int_{-\infty}^\infty d\omega dk\, V(x,t){\rm e}^{-i\omega t}\beta_k(\omega)\Psi_k(x).
\end{eqnarray}
Multiplying the last equation by ${\rm e}^{i\omega't}\Psi^*_k(x)$ and integrating with respect to $t$ and $x$, one obtains
\begin{eqnarray}
\label{eq4}
\hspace{-15mm}
\beta_{k'}(\omega')&=&\beta_{k'}^{(0)}(\omega')\nonumber \\
&+&\frac{\alpha}{4\pi^2\left(\hbar\omega'-\frac{\hbar^2k'^2}{2m}\right)}\int_{-\infty}^\infty d\omega dtdxdk\, V(x,t)\beta_k(\omega){\rm e}^{i(\omega'-\omega)t}\Psi^*_{k'}(x)\Psi_k(x)
\end{eqnarray}
with $\beta_{k'}^{(0)}(\omega')$ the expansion coefficient in the case $\alpha=0$.
In the last step, the orthogonality of the stationary eigenfunctions
\begin{equation}
\int_{-\infty}^\infty dx \Psi_{k'}^*(x) \Psi_{k}(x)=2\pi\delta(k-k')
\end{equation}
was used.
As Eq.\ (\ref{eq4}) is independent of $\alpha$ on the left hand side and the second term on the right hand side is linear in $\alpha$ it can be used to determine $\beta_k(\omega)$ in a recursive manner by inserting the solution for $\alpha=0$ on the right and thus calculating the solution linear in $\alpha$, inserting this again on the right to determine the solution quadratic in $\alpha$ and so on.

\subsection{Scattering induced by a fluctuating delta potential }\label{sect2}
We now  compute the current through a delta potential located at the position $\gamma(t)$ fluctuating randomly in time with the perturbation starting to act from 
time $t=0$ on. For further reference, we quote the transmission probability $|t_k|^2$ through a stationary potential of the form $\alpha \delta (x)$:

\begin{equation}
\label{eq:stationarydelta}
|t_k|^2=\frac{1}{1+\frac{\alpha^2m^2}{\hbar^4k^2}}=1-\frac{\alpha^2m^2}{\hbar^4k^2}+\ldots.
\end{equation}
expanded in the rightmost equation up to quadratic order in $\alpha$. We now analyze this quantity for the potential
\begin{equation}
\label{eq3000}
V(x,t)=\Theta(t)\delta(x-\gamma(t))\ ,
\end{equation}
where $\Theta(t)$ stands for the Heaviside function. The initial condition is the unperturbed solution
\begin{equation}
\hspace{-20mm}
\Psi_k(x)={\rm e}^{ikx}, \hspace*{0mm}\beta_k^{(0)}(\omega)=\delta\left(k-k_0\right)\delta\left(\omega-\frac{\hbar k_0^2}{2m}\right),\hspace*{0mm}\Psi^{(0)}(x,t)={\rm e}^{-i\hbar k_0^2t/(2m)}{\rm e}^{ik_0x}\ .
\end{equation}
Inserting this on the right hand side of Eq.\ (\ref{eq4}) yields
\begin{equation}
\label{eq5}
\beta_{k'}^{(1)}(\omega')=\frac{\alpha}{4\pi^2\left(\hbar\omega'-\frac{\hbar^2k'^2}{2m}\right)}\int_{0}^\infty dt\, {\rm e}^{i(\omega'-\hbar k_0^2/(2m))t}{\rm e}^{i\left(k_0-k'\right)\gamma(t)}
\end{equation}
and the correction linear in $\alpha$ is obtained from $\beta_{k'}^{(1)}(\omega')$ in Eq.\ (\ref{eq5}) by inserting it into Eq.\ (\ref{eq2})
\begin{eqnarray}
\label{eq6}
\fl\Psi^{(1)}(x,t)=\lim_{\epsilon\rightarrow0}\frac{\alpha}{4\pi^2\hbar}\int_{-\infty}^\infty d\omega'dk'\int_0^\infty dt'{\rm e}^{-i\omega't+ik'x} 
\frac{{\rm e}^{i\left(\omega'-\hbar k_0^2/(2m)\right)t'}}{\left(\omega'-\frac{\hbar k'^2}{2m}+i\epsilon\right)}{\rm e}^{i(k_0-k')\gamma(t')}.
\end{eqnarray}
Performing the $\omega'$-integral by the residual theorem, we get
\begin{eqnarray}
\label{eq6aa}
\fl\Psi^{(1)}(x,t)=\lim_{\epsilon\rightarrow0}\frac{\alpha}{2\pi i\hbar}\int_{-\infty}^\infty dk'\int_0^tdt'{\rm e}^{-i\hbar k't/(2m)+ik'x}{\rm e}^{i\hbar\left(k'^2-k_0^2\right)t'/(2m)}{\rm e}^{\epsilon t'}{\rm e}^{i(k_0-k')\gamma(t')}.
\end{eqnarray}
Finally we will not be interested in the current density for a specific realization of $\gamma(t)$ but in calculating the noise-averaged current density $j(x,t)$ perturbatively in $\alpha$ given by
\begin{equation}
j(x,t)=\left\langle\frac{\hbar}{2mi}\left[\Psi^*(x,t)\frac{\partial}{\partial x}\Psi(x,t)-c.c.\right]\right\rangle
\end{equation}
with $c.c.$ denoting the complex conjugate and where $\left\langle\ldots\right\rangle$ stands for the noise average. We will give details on how to perform it in the first
subsection of section \ref{sect5}. Starting from the unperturbed current density
\begin{equation}
j^{(0)}(x,t)=\left\langle\frac{\hbar}{2mi}\left[\Psi^{(0)*}(x,t)\frac{\partial}{\partial x}\Psi^{(0)}(x,t)-c.c.\right]\right\rangle=\frac{\hbar k_0}{m}
\end{equation}
the leading correction due to noise
in $\alpha$, $\Delta j^{(1)}(x,t)$, is given by
\begin{eqnarray}
\label{eq5000}
\hspace{-25mm}
\Delta j^{(1)}(x,t)=\left\langle\frac{\hbar}{2mi}\left[\Psi^{(0)*}(x,t)\frac{\partial}{\partial x}\Psi^{(1)}(x,t)+\Psi^{(1)*}(x,t)\frac{\partial}{\partial x}\Psi^{(0)}(x,t)-c.c.\right]\right\rangle.
\end{eqnarray}
As $\Psi^{(0)}(x,t)$ does not depend on the fluctuating potential, calculating this contribution is equivalent to considering the noise average of $\Psi^{(1)}(x,t)$ in Eq. (\ref{eq6}). 

As we will see below, $\Delta j^{(1)}(x,t)$ will vanish for this configuration. Therefore
we need to consider also the second order correction to the wave function. We insert the expression for $\beta_{k'}(\omega')$ from Eq.\ (\ref{eq5}) into the right hand side of Eq.\ (\ref{eq4}) yielding finally for $\Psi^{(2)}(x,t)$ by using (\ref{eq2})
\begin{eqnarray}
\label{eq9}
&\fl\Psi^{(2)}(x,t)=\lim_{\epsilon\rightarrow0}\frac{\alpha^2}{16\pi^4\hbar^2}\int_{-\infty}^\infty d\omega d\omega'dkdk'\int_0^\infty dt_1dt_2
{\rm e}^{i\omega\left(t_1-t_2\right)+i\omega'(t_2-t)}{\rm e}^{-i\hbar k_0^2t_1/(2m)}{\rm e}^{ik'x}\nonumber\\ &\fl\times\frac{1}{\left(\omega'-\frac{\hbar k'^2}{2m}+
i\epsilon\right)\left(\omega-\frac{\hbar k^2}{2m}+i\epsilon\right)}{\rm e}^{i\left(k_0-k\right)\gamma(t_1)}{\rm e}^{i\left(k-k'\right)\gamma(t_2)}\nonumber\\
&\fl=-\lim_{\epsilon\rightarrow0}\frac{\alpha^2}{4\pi^2\hbar^2}\int_{0}^tdt_2\int_{0}^{t_2}dt_1\int_{-\infty}^\infty dkdk'{\rm e}^{i\hbar k^2\left(t_1-t_2\right)/(2m)+
i\hbar k'^2(t_2-t)/(2m)}{\rm e}^{-i\hbar k_0^2t_1/(2m)}{\rm e}^{-\epsilon t_1}\nonumber\\&\fl\times{\rm e}^{ik'x}{\rm e}^{i\left(k_0-k\right)\gamma(t_1)}{\rm e}^{i
\left(k-k'\right)\gamma(t_2)}.
\end{eqnarray}
The noise averaged second order correction to the unperturbed current density
is then obtained as
\begin{eqnarray}
\label{eq1000}
\Delta j^{(2)}(x,t)&=&\frac{\hbar}{2mi}\left[\left\langle\Psi^{(1)*}(x,t)\frac{\partial}{\partial x}\Psi^{(1)}(x,t)\right\rangle+\Psi^{(0)*}(x,t)\frac{\partial}{\partial x}\left\langle\Psi^{(2)}(x,t)\right\rangle\right.\nonumber\\&&\left.+\left\langle\Psi^{(2)*}(x,t)\right\rangle\frac{\partial}{\partial x}\Psi^{(0)}(x,t)-c.c.\right].
\end{eqnarray}
In order to calculate $\Delta j^{(1)}(x,t)$ and $\Delta j^{(2)}(x,t)$ we now discuss how to perform the noise average.
\subsection{Noise average}
From Eqs.\ (\ref{eq6aa},\ref{eq9}) we obtain in $\Delta j^{(1)}(x,t)$ and $\Delta j^{(2)}(x,t)$ the noise dependent functions
\begin{equation}
\label{eq1001}
{\rm e}^{-ia\gamma(t_1)+ib\gamma(t_2)},
\end{equation}
where $a$ and $b$ are functions of $k_0$, $k$ and $k'$ in Eqs.\ (\ref{eq6aa},\ref{eq9}) and $t_1$ and $t_2$ are two time variables. We assume that the statistics of possible paths of $\gamma(t_1)$ and $\gamma(t_2)$ is described by an Ornstein-Uhlenbeck process  \cite{Uhlenbbeck, Wang+Uhl,Risken}, i.e.\ Brownian motion confined by a harmonic force.
The advantage of this process compared to the Wiener process \cite{Risken,Leschke}  is that the variance of $\gamma(t_i)$ remains bounded also for large $t_i$
\begin{equation}
\label{var}
\left\langle\gamma^2(t_i)\right\rangle=\frac{\sigma^2}{2\theta}\left(1-{\rm e}^{-2\theta t_i}\right)
\end{equation}
with the parameter $\theta$ characterizing the time necessary to obtain a stationary distribution and the parameter $\sigma$ which is determined by the strength of the 
harmonic potential whereas we have for a Wiener process
\begin{equation}
\left\langle\gamma^2(t_i)\right\rangle\propto t.
\end{equation}
We can thus assume that the perturbation remains localized for the Ornstein Uhlenbeck process within a certain region, e.g.\ one bond of a graph.
 For an Ornstein Uhlenbeck process, the transition probability $P(\gamma_2,t_2;\gamma_1,t_1)$ from position $\gamma_1$ and time $t_1$ to position $\gamma_2$ and time $t_2$ is given by \cite{Risken}
\begin{equation}
\hspace{-20mm}
P(\gamma_2,t_2;\gamma_1,t_1)=\sqrt{\frac{\theta}{2\pi\sigma^2\left(1-{\rm e}^{-2\theta(t_2-t_1)}\right)}}\exp\left[-\frac{\theta\left(\gamma_2-
{\rm e}^{-\theta(t_2-t_1)}\gamma_1\right)^2}{2\sigma^2\left(1-{\rm e}^{-2\theta(t_2-t_1)}\right)}\right],
\end{equation}
that yields for large $\theta(t_2-t_1)$
\begin{equation}
\label{measure}
P(\gamma_2,t_2;\gamma_1,t_1)=\sqrt{\frac{\theta}{2\pi\sigma^2}}\exp\left[-\frac{\theta \gamma_2^2}{2\sigma^2}\right].
\end{equation}
In this limit  the average of Eq.\ (\ref{eq1001}) is obtained by integrating it with a Gaussian measure according to Eq.\ (\ref{measure}):
\begin{equation}
\label{aver}
\left\langle{\rm e}^{-ia\gamma(t_1)+ib\gamma(t_2)}\right\rangle={\rm e}^{-(a^2+b^2)\sigma^2/(2\theta)}.
\end{equation}
The regime of $\sigma^2/(2\theta)$-values where this analysis is valid is restricted by Eq.\ (\ref{var}): In case the interval where the delta-potential is allowed to move possesses length $L$, this can be regarded in a graph as the length of the bond where the perturber is placed, we obtain
\begin{equation}
\label{rest}
\frac{\sigma^2}{2\theta}\ll L^2.
\end{equation}
In the next subsection we use Eq.\ (\ref{aver}) to calculate the average contributions to the current density.

\subsection{Noise averaged current correction}\label{sections4}

Using (\ref{aver}) we calculate the noise averaged quantities appearing in the current corrections in Eqs.\ (\ref{eq5000},\ref{eq1000}):
$\left\langle\Psi^{(1)}(x,t)\right\rangle$, $\left\langle\Psi^{*(1)}(x,t)\partial_x\Psi^{(1)}(x,t)\right\rangle$ and $\left\langle\Psi^{(2)}(x,t)\right\rangle$.
We give details of this calculation in section \ref{sect5}.

For $\Psi^{(0)*}(x,t)\left\langle\Psi^{(1)}(x,t)\right\rangle$ we get a real contribution as shown before Eq.\ (\ref{firsto}) leading to $\Delta j^{(1)}(x,t)=0$ fir the configuration
considered here.

Next we consider $\Delta j^{(2)}(x,t)$: Here we need to add the contributions from $\left\langle\Psi^{(1)*}(x,t)\partial_x\Psi^{(1)}(x,t)\right\rangle$ in 
Eqs.\ (\ref{firsto},\ref{split}) and from $\left\langle\Psi^{(2)}(x,t)\right\rangle$ in Eq.\ (\ref{eq1010}). We obtain 
in total for $\Delta j^{(2)}(x,t)$ defined  in Eq.\ (\ref{eq1000})
\begin{equation}
\label{eq7000}
\Delta j^{(2)}(x,t)=-\frac{\alpha^2m}{\hbar^3k_0}{\rm e}^{-2\sigma^2k_0^2/\theta}.
\end{equation}
Note that for $\sigma=0$ this result agrees with the correction obtained for a static delta potential (\ref {eq:stationarydelta}) up to quadratic order in $\alpha$. Increasing the variance of the fluctuations, i.e.\ increasing $\sigma^2/\theta$, the effect of the delta potential on the current is reduced compared to the static setup. We obtain for $\sigma^2/\theta=\ln2/(2k_0^2)$ only half of the current compared to $\sigma^2=0$. This implies that the current decay is for fixed $\sigma^2/\theta$ especially pronounced in the regime of large $k_0^2$, i.e.\ in the regime of large energies of the incoming wave.

This can be understood by noting that this current decrease results from the additional phase ${\rm e}^{i(k_0-k)\gamma(t_1)-i(k'-k)\gamma(t_2)}$ that the second order correction to the wave function acquires in Eq.\ (\ref{eq9}). The residual integrations set $k'=k_0$ and $k=\pm k_0$. For $\gamma(t_1)\neq\gamma(t_2)$ we obtain an additional phase in the wave function for $k=-k_0$ that reduces its contribution to the current density. This phase fluctuates stronger with increasing $k_0$ inducing a stronger suppression of this current density contribution.
The computation for the effect of the noisy perturbation on the transmission through an arbitrary quantum  graph use essentially the same tools as described above. However, the existence of resonances in the transmission spectrum of the unperturbed spectrum leads to richer effects.

\section{Time dependent perturbations on  graphs}\label{sect3}
We shall now consider  an open quantum graph where the set of bonds consists of $l$ leads connecting vertices to the outside and $B$ inner bonds connecting between vertices as prescribed by the graph adjacency matrix.  Random noise is  introduced by a $\delta$-potential on one of the interior bonds with its position fluctuating with time. One can also consider this as an extra vertex placed on the bond at a time dependent position. The noise averaged current density through the graph will be computed to second order in the coupling strength.
\subsection{General formalism}
The stationary wave functions at wave number $k_0$ in the absence of a time dependent potential  consist of incoming and outgoing waves on the leads,
\begin{equation}
\label{eq12a}
\Psi_i(x)=C^{\rm in}_i(k_0){\rm e}^{-ik_0x}+C^{\rm out}_i(k_0){\rm e}^{ik_0x},\ \  {\rm for\ all\ } 1\le i\le l\ .
\end{equation}
The wave function on the interior bonds can again be written as a superposition of counter propagating waves:
\begin{equation}
\label{eq14}
\Psi_j(x)=A_j(k_0) {\rm e}^{ik_0x}+B_j(k_0){\rm e}^{-ik_0x}\ \  {\rm for\ all\ } 1\le j\le B\ .
\end{equation}
The coefficients $C^{\rm in}_i(k_0)$, $C^{\rm out}_i(k_0)$, $A_j(k_0)$ and $B_j(k_0)$ can be determined \cite{Band} for any set of boundary conditions which renders 
the problem self adjoint \cite{BerkoKuch}. The boundary conditions imposed here require  that the wave functions are continuous at the vertices
\begin{equation}
\label{eq15}
\Psi_e(v)=\Psi_{e'}(v)
\end{equation}
for all bonds $e$ and $e'$ reaching vertex $v$. Additionally the outgoing derivatives of all the wave functions at each vertex sum to a certain value $\alpha_v\Psi(v)$, implying
\begin{equation}
\label{eq16}
\sum_{e=1}^d\frac{d\Psi_e}{dx_e}(v)=\alpha_v\Psi(v)
\end{equation}
for a vertex connected to $d$ bonds.

The noise inducing time dependent $\delta$-potential (\ref{eq3000}) is now attached to one interior bond at a mean position $x_0$ and the time dependent fluctuations given by  $\gamma(t)$. This function should obey the condition $0<x_0+\gamma(t)< L$ with $L$ the length of the bond where the perturbation is placed, this condition is again translated into the one given in Eq.\ (\ref{rest}) when we will later consider the dynamics of $\gamma(t)$ to be described by an Ornstein Uhlenbeck process.

Compared to the calculation in the last section we need to insert the unperturbed scattering solutions to obtain the first two order corrections in $\alpha$ instead of the 
plane wave solutions.
To be more precise,  for any fixed $k_0$ there are  $l$ scattering states with the $i$-th state taking on lead $j$ the form in a coordinate system pointing towards the scattering region
\begin{equation}
\Psi_{i,k_0}^j(x)=\delta_{ij}{\rm e}^{ik_0x}+S_{ji}(k_0){\rm e}^{-ik_0x},
\end{equation}
i.e.\ there is an incoming wave only on lead $i$ and outgoing waves on all leads.
These functions are  complete and orthogonal \cite{Scatt} as required in Eq.\ (\ref{eq2}) in the sense
\begin{eqnarray}
\label{orth}
\int_\mathcal{G} dx\Psi^*_{i,k_0}(x)\Psi_{o,k_0'}(x)=2\pi\delta_{io}\delta(k_0-k_0')\nonumber\\
\sum_{i=1}^l\int_0^\infty dk_0\Psi_{i,k_0}(x)\Psi_{i,k_0}^{*}(x')=2\pi\delta(x-x')
\end{eqnarray}
with $\mathcal{G}$ denoting the whole graph. The fact that the conditions (\ref{orth}) are fulfilled makes the relations derived in  section \ref{sect1}  applicable. The main difficulty in using these data in the formulation of section  \ref{sect1}  is the $k_0$ dependence of the coefficients $A_j,B_j,C_i^{{\rm in/out}}$ (\ref{eq12a},\ref{eq14}). The $k_0$ dependence of these coefficients reflects the presence of resonances and renders the required integrations very complicated (albeit numerically accessible). In order to illustrate the effect of the noise in the presence of resonances we shall analyze a simple yet not trivial system which can be addressed analytically, and will defer the discussion of the general situation to the last part of this section. The system to be used is described  in the next subsection.

\begin{figure}
\begin{center}
\includegraphics[width=8cm]{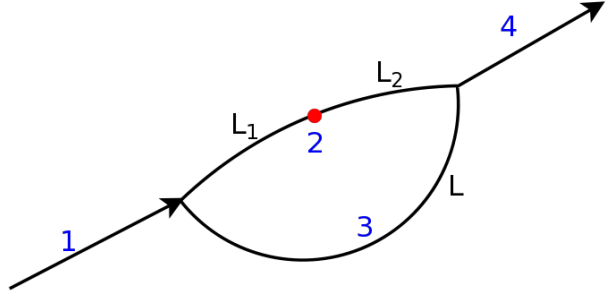}
\end{center}
\caption{The loop graph. The dot on bond 2 indicates where the $ \delta$ potential is to be added}
\label{fig1}
\end{figure}

\subsection {Scattering from a Loop}
The loop graph is shown in Fig.\ \ref{fig1}, with two bonds (2,3) joining to make a loop, and two leads (1,4) emerging from them. The length of bond 2 will be 
denoted by $\hat L=L_1+L_2$ and that of bond 3 will be denoted by $L$. (The time averaged position of the $\delta$ interaction will be on the bond 2 at a distance 
$L_1$ from the left vertex, and $L_2 = {\hat L} -L_1 $ away from the right vertex).

\begin{figure}
\begin{center}
\includegraphics[width=8cm]{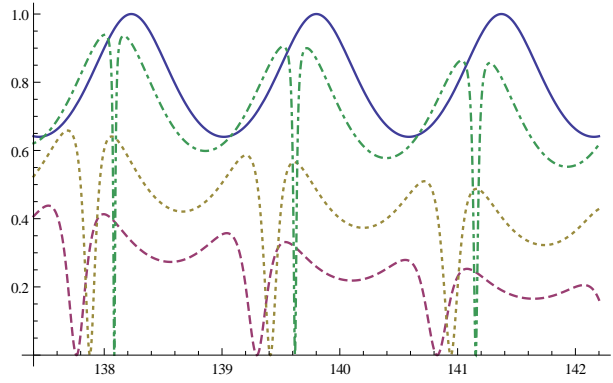}
\caption{Transmission through the graph in Fig.\ \ref{fig1} for different ratios of $L$ and $\hat{L}=L_1+L_2$.}
\label{figu}
\end{center}
\end{figure}

The components of the scattering  solutions for the unperturbed loop on the different bonds read:
\begin{eqnarray}
1:&&\Psi(x)={\rm e}^{ik_0x}+R(k_0){\rm e}^{-ik_0x}\nonumber\\
2:&&\Psi(x)=A(k_0){\rm e}^{ik_0x}+B(k_0){\rm e}^{-ik_0x}\nonumber\\
3:&&\Psi(x)=C(k_0){\rm e}^{ik_0x}+D(k_0){\rm e}^{-ik_0x}\nonumber\\
4:&&\Psi(x)=T(k_0){\rm e}^{ik_0x},
\end{eqnarray}
i.e.\ an incoming wave only on the left lead.
From the conditions (\ref{eq15}) and (\ref{eq16}) with $\alpha_v=0$ we get the following results for $T(k_0)$, $A(k_0)$, $B(k_0)$, $C(k_0)$, $D(k_0)$ and $R(k_0)$ 
\begin{eqnarray}
\label{eq19}
T(k_0)&=&\frac{8i\left[{\sin}\,k_0L+{\sin}\,k_0{\hat L}\right]}{\mathcal{D}(k_0)},\nonumber\\
A(k_0)&=&\frac{2\left[{\rm e}^{ik_0(L-{\hat L})}+2-3{\rm e}^{-ik_0(L+{\hat L})}\right]}{\mathcal{D}(k_0)},\nonumber\\
B(k_0)&=&\frac{2\left[{\rm e}^{ik_0(L+{\hat L})}-2+ {\rm e}^{-ik_0(L-{\hat L})}\right]}{\mathcal{D}(k_0)},\nonumber\\
C(k_0)&=&\frac{2\left[2+{\rm e}^{-ik_0(L-{\hat L})}-3{\rm e}^{-ik_0(L+{\hat L})}\right]}{\mathcal{D}(k_0)},\nonumber\\
D(k_0)&=&\frac{2\left[{\rm e}^{ik_0(L+{\hat L})}-2+{\rm e}^{ik_0(L-{\hat L})}\right]}{\mathcal{D}(k_0)},\nonumber\\
R(k_0)&=&\frac{6{\cos}\,k_0(L+{\hat L})-8+2{\cos}\,k_0(L-{\hat L})}{\mathcal{D}(k_0)}
\end{eqnarray}
with
\begin{equation}\label{detr}
\mathcal{D}(k_0)\equiv 8-{\rm e}^{ik_0(L+{\hat L})}+{\rm e}^{ik_0(L-{\hat L})}+ {\rm e}^{ik_0({\hat L}-L)}-9{\rm e}^{-ik_0(L+{\hat L})}.
\end{equation}
 The wave function is a meromorphic function of $k_0$ with poles at the complex zeros of $\mathcal{D}(k_0)$. These poles induce the resonances in this scattering 
 system, and their distance from the real axis determines the resonance widths. It was shown in \cite {Waltner} that arbitrarily narrow resonances can be found when 
 the ratio of the lengths of the two parts of the loop differ slightly from integer values. These are the "topological resonances" in the present system. At rational 
 values of ${\hat L}/L$ the resonances which dominate the scattering process are the broad "shape resonances" which result from the poles which are away from the 
 real axis. The two cases are illustrated in Fig. \ref{figu}, where we plot the transmission from the left to the right lead in the range $kL_0\in\left[137.4,142.1\right]$ for $\hat{L}=2L_0$ and $L=(2+\delta)L_0$ with $\delta=0$
 for the full (blue) curve, $\delta=\pi/30$ for the dashed (violet), $\delta=\pi/31$ for the dotted (brown) and $\delta=\pi/33$ for the dashed dotted (green) curve.  

We shall now return to the main issue of the paper and show how the introduction of noise affects the two kinds of resonances which appear in scattering from the loop.

\subsection {Scattering from a noisy loop}

The noise is introduced by adding a $\delta$ potential on bond "2". Its position is fluctuating in time, but centered at a distance $L_1$ from the left vertex, as indicated by the dot in Fig \ref{fig1}.
Assuming an incident wave with wave number $k_0$ coming from the left lead, we use the expressions derived in the previous section to derive the mean current through the loop to second order in $\alpha$. The stationary solution on bond "2" is given by
\begin{equation}\label{wave}
\Psi(x)=A(k_0){\rm e}^{ik_0x}+B(k_0){\rm e}^{-ik_0x}.
\end{equation}
With this we get instead of Eq.\ (\ref{eq6}) for the first order correction to the wave function on right lead "4" in Fig.\ \ref{fig1}
\begin{eqnarray}
\label{eq1030}
&\fl\Psi^{(1)}(x,t)=\frac{-i\alpha}{2\pi\hbar}\lim_{\epsilon\rightarrow 0}\int_{0}^t dt'\int_0^\infty dk'{\rm e}^{-i\hbar k'^2t/(2m)}\left[A(k_0){\rm e}^{ik_0(L_1+\gamma(t'))}+B(k_0){\rm e}^{-ik_0(L_1+\gamma(t'))}\right] \nonumber\\&\fl\times 
{\rm e}^{i\left(\hbar k'^2/(2m)-\hbar k_0^2/(2m)-i\epsilon\right)t'}\left\{C(k'){\rm e}^{ik'x}\left[A^*(k'){\rm e}^{-ik'(L_1+\gamma(t'))}+B^*(k'){\rm e}^{ik'(L_1+\gamma(t'))}\right]\right.\nonumber\\&\fl\left.
+\left[R(k'){\rm e}^{ik'x}+{\rm e}^{-ik'x}\right]\left[A^*(k'){\rm e}^{-ik'(L_2-\gamma(t'))}+B^*(k'){\rm e}^{ik'(L_2-\gamma(t'))}\right]\right\},
\end{eqnarray}
where the term proportional to $C(k')$ results from the scattering state with an incoming wave on the left lead and the other term from a scattering eigenstate with an 
incoming wave on the right lead. Remember that according to Eq.\ (\ref{orth}) we only have a complete orthogonal basis when considering incoming waves on every lead. 
Due to the left right symmetry of the considered graph in the latter case the wave function where the additional delta potential is placed takes the same form as in Eq.\ 
(\ref{wave}) however now in a coordinate system pointing leftwards instead of rightwards with origin at the right vertex instead of at the left vertex. The perturbing delta potential is in this coordinate system placed at $L_2-\gamma(t)$. To generalize Eq.\ (\ref{eq1030}) to an arbitrary graph we need to add the contributions from all scattering eigenstates instead of two as introduced above and consider the corresponding wave functions where the perturbation is placed.
For the second order correction we get instead of Eq.\ (\ref{eq9}) in an analogous way as in Eq.\ (\ref{eq1030})
\begin{eqnarray}
\label{eq1031}
&\fl\Psi^{(2)}(x,t)=-\frac{\alpha^2}{4\pi^2\hbar^2}\lim_{\epsilon\rightarrow 0}\int_{0}^t dt_2\int_0^{t_2}\!\!dt_1\!\int_0^\infty\!\!\!\!\!\! dk_1dk_2
{\rm e}^{i\hbar k_1^2\left(t_1-t_2\right)/(2m)+i\hbar k_2^2(t_2-t)/(2m)-i\hbar k_0^2t_1/(2m)-\epsilon t_1}\nonumber\\&\fl\times\left\{C(k_2){\rm e}^{ik_2x}\left[A^*(k_2)
{\rm e}^{-ik_2(L_1+\gamma(t_2))}+B^*(k_2){\rm e}^{ik_2(L_1+\gamma(t_2))}\right]+\left[R(k_2){\rm e}^{ik_2x}+{\rm e}^{-ik_2x}\right]\times\right.\nonumber\\&\fl\left.\left[A^*(k_2)
{\rm e}^{-ik_2(L_2-\gamma(t_2))}+B^*(k_2){\rm e}^{ik_2(L_2-\gamma(t_2))}\right]\right\}\!\!\!\left[A(k_0){\rm e}^{ik_0(L_1+\gamma(t_1))}+B(k_0)
{\rm e}^{-ik_0(L_1+\gamma(t_1))}\right]\times\nonumber\\&\fl\left\{\left[A^*(k_1){\rm e}^{-ik_1(L_1+\gamma(t_1))}+B^*(k_1){\rm e}^{ik_1(L_1+\gamma(t_1))}\right]\!\!\!\left[A(k_1)
{\rm e}^{ik_1(L_1+\gamma(t_2))}+B(k_1){\rm e}^{-ik_1(L_1+\gamma(t_2))}\right]+\right.\nonumber\\&\fl\left.\left[A^*(k_1){\rm e}^{-ik_1(L_2-\gamma(t_1))}+B^*(k_1)
{\rm e}^{ik_1(L_2-\gamma(t_1))}\right]\left[A(k_1){\rm e}^{ik_1(L_2-\gamma(t_2))}+B(k_1){\rm e}^{-ik_1(L_2-\gamma(t_2))}\right]\right\}.\nonumber\\
\end{eqnarray}
Also here the expression must  include  all contributions from scattering eigenstates and the corresponding wave functions at the place of the perturber in the case of a general graph.

Finally we again are not interested in the corrections to the wave functions for a specific $\gamma(t)$ but its average behavior when the dynamics of $\gamma(t)$ is described by an Ornstein Uhlenbeck process.
Technically the expressions for $\Psi^{(1)}(x,t)$ and $\Psi^{(2)}(x,t)$ for a graph induce two changes compared  to before when averaging that we explain in detail in section \ref{sect5}. First, we get due to the fact that each stationary eigenfunction consists of two counter propagating waves several contributions. Second, the coefficients $C(k)$, $R(k)$, $A(k)$ and $B(k)$ possess poles in the complex $k$ plane leading to additional contributions when performing the integrations with respect to the wavenumbers.

After performing these averages according to Eq.\ (\ref{aver}) and the remaining integrals in Eq.\ (\ref{eq1030}) we get the first order correction  $\left\langle\Psi^{(1)}(x,t)\right\rangle$
as calculated in detail in Eq.\ (\ref{eq1032}). This results in the following contribution to the current density linear in $\alpha$, Eq.\ (\ref{eq5000}),
\begin{eqnarray}\label{cur1}
&\fl\left\langle j^{(1)}(x,t)\right\rangle=\frac{2\alpha }{\hbar}\Im\left\{C^*(k_0)R(k_0)\left[A^*(k_0)B(k_0){\rm e}^{-ik_0(L_2+L_1)}+B^*(k_0)A(k_0){\rm e}^{ik_0(L_2+L_1)}
\right.\right.\nonumber\\&\fl\left.\left.+{\rm e}^{-2\sigma^2k_0^2/\theta}\left(|A(k_0)|^2{\rm e}^{-ik_0(L_2-L_1)}+|B(k_0)|^2{\rm e}^{ik_0(L_2-L_1)}\right)\right]\right\}.\nonumber
\end{eqnarray}
Next the contribution $\Delta j^{(2)}(x,t)$ defined in Eq.\ (\ref{eq1000}) is evaluated, we get
\begin{equation}\label{cur2}
\fl\Delta j^{(2)}(x,t)=\frac{\hbar k_0}{m}\Im\left[i\left\langle\Psi^{(1)*}(x,t)\right\rangle\left\langle\Psi^{(1)}(x,t)\right\rangle+2i\Psi^{(0)*}(x,t)
\left\langle\Psi^{(2)}(x,t)\right\rangle\right].
\end{equation}
We need therefore the contributions $\left\langle\Psi^{(1)*}(x,t)\right\rangle\left\langle\Psi^{(1)}(x,t)\right\rangle$ obtained from Eqs.\ (\ref{split},\ref{eq1032})
\begin{eqnarray}
&\fl\left\langle\Psi^{(1)*}(x,t)\right\rangle\left\langle\Psi^{(1)}(x,t)\right\rangle=\frac{\alpha^2m^2}{\hbar^4k_0^2}
\left|\left\{C(k_0)\left[|A(k_0)|^2+|B(k_0)|^2+{\rm e}^{-2\sigma^2k_0^2/\theta}\right.\right.\right.\nonumber\\&\fl\left.\left.\left.
\times\left(A(k_0)B^*(k_0){\rm e}^{2ik_0L_1}+A^*(k_0)B(k_0){\rm e}^{-2ik_0L_1}\right)\right]+R(k_0)\left[A(k_0)B^*(k_0){\rm e}^{ik_0(L_2+L_1)}+\right.\right.\right.
\nonumber\\&\fl\left.\left.\left.\times A^*(k_0)B(k_0){\rm e}^{-ik_0(L_2+L_1)}+{\rm e}^{-2\sigma^2k_0^2/\theta}\left(|A(k_0)|^2{\rm e}^{-ik_0(L_2-L_1)}+|B(k_0)|^2{\rm e}^{ik_0(L_2-L_1)}\right)\right]\right\}\right|^2
\end{eqnarray}
and the ones resulting from $\left\langle\Psi^{(2)}(x,t)\right\rangle$ given in Eqs.\ (\ref{eq1037}-\ref{letz}). We split this contribution into three parts, the 
first one results from all terms proportional to $C(k_0)$ in Eq.\ (\ref{eq1031}) indexed by $C$
\begin{eqnarray}
&\fl\Re\left[\Psi^{(0)*}(x,t)\left\langle\Psi^{(2)}(x,t)\right\rangle\right]_C=-\frac{\alpha^2m^2}{2\hbar^4k_0^2}|C(k_0)|^2
\left\{\left[\left(1+{\rm e}^{-4\sigma^2k_0^2/\theta}\right)\left(|A(k_0)|^2+|B(k_0)|^2\right)^2\right.\right.\nonumber\\&\fl\left.\left.
+2\Re\left[\left(A^*(k_0)B(k_0){\rm e}^{-2ik_0L_1}{\rm e}^{-4\sigma^2k_0^2/\theta}+A(k_0)B^*(k_0){\rm e}^{2ik_0L_1}\right)\left(A(k_0)B^*(k_0)
{\rm e}^{2ik_0L_2}\right.
\right.\right.\right.\nonumber\\&\fl\left.\left.\left.\left.+A^*(k_0)B(k_0){\rm e}^{-2ik_0L_1}\right)\right]\right]+2{\rm e}^{-2\sigma^2k_0^2/\theta}\left(|A(k_0)|^2+|B(k_0)|^2\right)\nonumber
\right.\\&\fl\left.\times\Re\left[\left(3A(k_0)B^*(k_0){\rm e}^{2ik_0L_1}+A(k_0)B^*(k_0){\rm e}^{2ik_0L_2}\right)\right]\right\}
\end{eqnarray}
the next one from terms proportional to $R(k_0)$ in Eq.\ (\ref{eq1031}) obtained when considering the poles at $k_1=\pm k_0$ in the $k_1$-integration
\begin{eqnarray}
&\fl\Re\left[\Psi^{(0)*}(x,t)\left\langle\Psi^{(2)}(x,t)\right\rangle\right]_R=
-\frac{\alpha^2m^2}{2\hbar^4k_0^2}\Re\left\{C^*(k_0)R(k_0)\left[2{\rm e}^{-2\sigma^2k_0^2/\theta}\left(1+{\rm Erf}(i\sigma k_0/\sqrt{\theta})\right)\right.\right. 
\nonumber\\ &\fl\left.\left.\times\left(\left(|A(k_0)|^4+|A(k_0)|^2|B(k_0)|^2\right){\rm e}^{-ik_0(L_2-L_1)}+
\left(|B(k_0)|^4+|A(k_0)|^2|B(k_0)|^2\right){\rm e}^{ik_0(L_2-L_1)}\right.\right.\right.\nonumber\\ &\fl\left.\left.\left.+4\Re\left(A(k_0)B^*(k_0){\rm e}^{ik_0(L_2+L_1)}\right)\Re\left(A^*(k_0)B(k_0)\left({\rm e}^{-2ik_0L_2}+{\rm e}^{-2ik_0L_1}\right)\right)\right)\right.\right.\nonumber\\& \fl
\left.\left.
+4\left(|A(k_0)|^2+|B(k_0)|^2\right)\left[\Re
\left(A(k_0)B^*(k_0){\rm e}^{ik_0(L_2+L_1)}\right)+\frac{1}{2}{\rm e}^{-4\sigma^2k_0^2/\theta}\right.\right.\right.\nonumber\\&\fl\left.\left.\left.\times\left(1+{\rm Erf}(2i\sigma k_0/\sqrt{\theta})\right)\Re\left(A(k_0)B^*(k_0){\rm e}^{ik_0(L_2+L_1)}\right)\right]+{\rm e}^{-4\sigma^2 k_0^2/\theta}\left(1+{\rm Erf}
(2i\sigma k_0/\sqrt{\theta})\right)
\right.\right.\nonumber\\&\fl\left.\left.\times
\left[|A(k_0)|^2{\rm e}^{-ik_0(L_2-L_1)}\left(A^*(k_0)B(k_0){\rm e}^{-ik_0(L_2+L_1)}+A(k_0)B^*(k_0){\rm e}^{2ik_0L_1}\right)+
|B(k_0)|^2\right.\right.\right.\nonumber\\&\fl\left.\left.\left.\times{\rm e}^{ik_0(L_2-L_1)}\left(A^*(k_0)B(k_0){\rm e}^{-2ik_0L_2}+A(k_0)B^*(k_0){\rm e}^{2ik_0L_1}\right)\right]
\right]\right\}
\end{eqnarray}
and the final one resulting from terms proportional to $R(k_0)$ in Eq.\ (\ref{eq1031}) obtained when considering the potential poles at $\tilde{k}_1$ and $-\tilde{k}_1^*
$ of $A(k_1)$, where $\tilde{k}_1$ is assumed to be situated in the upper right complex plane
\begin{eqnarray}\label{cur5}
&\fl\Re\left[\Psi^{(0)*}(x,t)\left\langle\Psi^{(2)}(x,t)\right\rangle\right]_P=
\frac{\alpha^2m^2}{\hbar^4k_0^2}\Im\left\{C^*(k_0)R(k_0)8\Im\left[\frac{k_0}{\tilde{k}_1^2-k_0^2}
{\rm e}^{-\sigma^2\left(\tilde{k}_1^2+k_0^2\right)/\theta}\right.\right.\nonumber\\&\fl\left.\left.\times\left(1+{\rm Erf}(i\sigma\tilde{k}_1/\sqrt{\theta})\right)
\left[\left(|A(k_0)|^2{\rm e}^{-ik_0(L_2-L_1)}+|B(k_0)|^2{\rm e}^{ik_0(L_2-L_1)}\right)\right.\right.\right.\nonumber\\&\fl\left.\left.\left.\times
\left(A^*(\tilde{k}_1)\frac{A_o(\tilde{k}_1)}{\mathcal{D}_u'(\tilde{k}_1)}+B^*(\tilde{k}_1)\frac{B_o(\tilde{k}_1)}{\mathcal{D}_u'(\tilde{k}_1)}
\right)+\Re\left(A^*(k_0)B(k_0){\rm e}^{-ik_0(L_2+L_1)}\right)\left[B^*(\tilde{k}_1)\frac{A_o(\tilde{k}_1)}{\mathcal{D}_u'(\tilde{k}_1)}\right.\right.\right.\right.\nonumber\\&\fl\left.\left.\left.\left.\times\left({\rm e}^{2i\tilde{k}_1L_1}+{\rm e}^{2i\tilde{k}_1L_2}\right)+A^*(\tilde{k}_1)\frac{B_o(\tilde{k}_1)}{\mathcal{D}_u'(\tilde{k}_1)}\left({\rm e}^{-2i\tilde{k}_1L_1}+{\rm e}^{-2i\tilde{k}_1L_2}\right)\right]\right]\right]+4\Im\left[\frac{k_0}{\tilde{k}_1^2-k_0^2}\right.\right.
\nonumber\\&\fl\left.\left.\times\left[{\rm e}^{-\sigma^2(\tilde{k}_1-k_0)^2/\theta}\left(1+{\rm Erf}\left(i\sigma\left(\tilde{k}_1-k_0\right)\!/\!\sqrt{\theta}\right)\right)
+{\rm e}^{-\sigma^2(\tilde{k}_1+k_0)^2/\theta}\left({\rm Erf}\left(i\sigma\left(\tilde{k}_1+k_0\right)\!/\!\sqrt{\theta}\right)\right.\right.\right.\right.
\nonumber\\&\fl\left.\left.\left.\left.+1\right)\right]2\Re\left(A(k_0)B^*(k_0){\rm e}^{ik_0(L_2+L_1)}\right)\left(A^*(\tilde{k}_1)\frac{A_o(\tilde{k}_1)}{\mathcal{D}_u'(\tilde{k}_1)}+B^*(\tilde{k}_1)\frac{B_o(\tilde{k}_1)}
{\mathcal{D}_u'(\tilde{k}_1)}\right)+\frac{k_0}{\tilde{k}_1^2-k_0^2}\right.\right.\nonumber\\&\fl\left.\left.\times
\left[{\rm e}^{-\sigma^2(\tilde{k}_1-k_0)^2/\theta}\left(1+{\rm Erf}(i\sigma(\tilde{k}_1-k_0)/\sqrt{\theta})\right)|A(k_0)|^2{\rm e}^{-i(k_0-\tilde{k}_1)(L_2-L_1)}\right.\right.\right.\nonumber\\&\fl\left.\left.\left.
+{\rm e}^{-\sigma^2(\tilde{k}_1+k_0)^2/\theta}\left(1+{\rm Erf}(i\sigma(\tilde{k}_1+k_0)/\sqrt{\theta})\right)|B(k_0)|^2{\rm e }^{i(k_0-\tilde{k}_1)(L_2-L_1)}\right]\right.\right.
\nonumber\\&\fl\left.\left.\times\left(A^*(\tilde{k}_1)\frac{B_o(\tilde{k}_1)}{\mathcal{D}_u'(\tilde{k}_1)}{\rm e}^{-i\tilde{k}_1(L_2+L_1)}+B^*(\tilde{k}_1)\frac{A_o(\tilde{k}_1)}
{\mathcal{D}_u'(\tilde{k}_1)}{\rm e}^{i\tilde{k}_1(L_2+L_1)}\right)\right]\right\}
\end{eqnarray}
We introduced here the notation $A(k)\equiv A_o(k)/\mathcal{D}(k)$ and $B(k)\equiv B_o(k)/\mathcal{D}(k)$ with  $A_o(k)$ and $B_o(k)$ the functions in the numerator 
of $A(k)$ and $B(k)$ in Eq.\ (\ref{eq19}), respectively and $\mathcal{D}(k)$ defined in Eq.\ (\ref{detr}). Note that also the functions $A^*(k_1)$ and $B^*(k_1)$ possess
poles at $-\tilde{k}_1$ and $\tilde{k}_1^*$ as its denominator is complex conjugated compared to the one of $A(k_1)$ and $B(k_1)$. In the case of several zeros of $\mathcal{D}(k_1)$
we need to sum the contribution (\ref{cur5}) resulting from the different poles.

Before coming to a numerical illustration of these expressions for the current density we comment on how to make this formalism applicable to a general graph: In a first step 
we need to determine $\Psi_1(x,t)$ and $\Psi_2(x,t)$, therefore we calculate all scattering eigenstates and the form of the wave function where the fluctuating delta potential
is placed using the conditions (\ref{eq15}) and (\ref{eq16}).  In a next step the functions of $\Psi_1(x,t)$ and $\Psi_2(x,t)$ relevant for the current density corrections in Eqs.\ (\ref{eq5000},\ref{eq1000}) need to be averaged
with respect to the noise. From this procedure we get the current correction resulting from the noise in terms of the coefficients of the stationary wave functions on the graph. These
complex expressions can be analyzed numerically.

\section{Numerical illustrations}\label{sect4}
\begin{figure}
\begin{center}
\includegraphics[width=12cm]{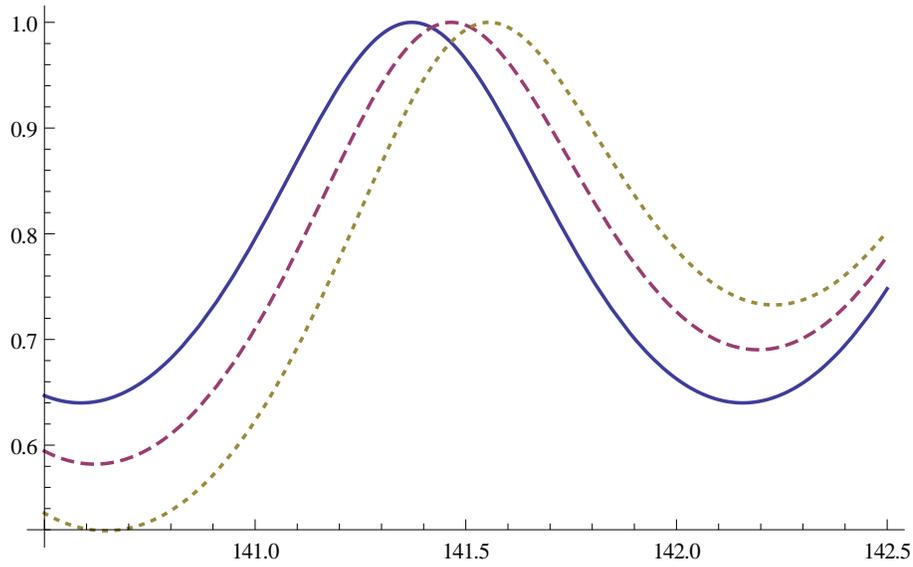}
\caption{The $k$-dependence of the transmission through  a loop due to a {\it static} perturbation, for $kL_0\in[140.5,142.5]$ - the vicinity of a shape resonance.  
Three cases are shown: unperturbed transmission $m\alpha/(\hbar^2k_0)=0$  (blue full line), $m\alpha/(\hbar^2k_0)=20/(k_0L_0)$ (violet dashed line), $m\alpha/(\hbar^2
k_0)=40/(k_0L_0)$ (brown dotted line).}
\label{dia1}
\end{center}
\end{figure}

\begin{figure}
\begin{center}
\includegraphics[width=12cm]{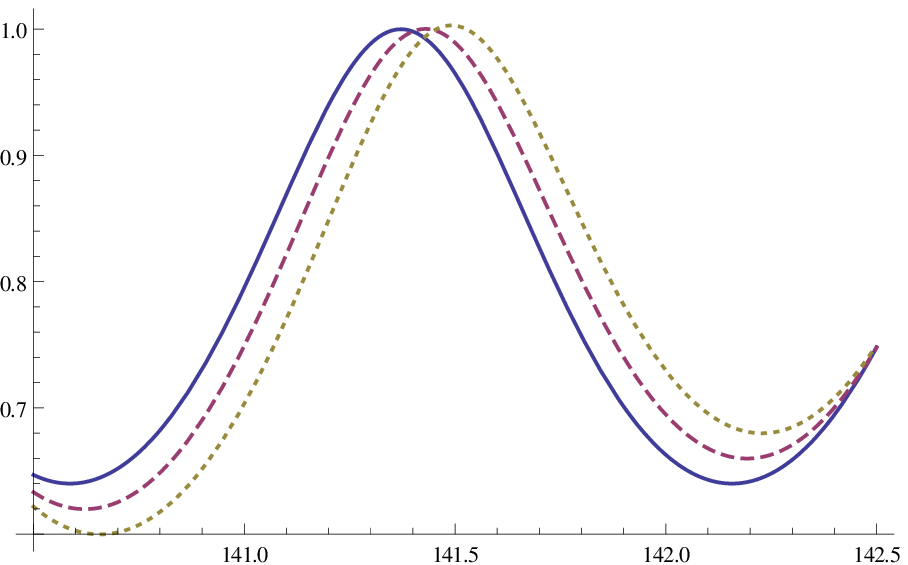}
\caption{The effect of noise characterized by  $\sigma k_0/\sqrt{\theta}=0.005 k_0L_0 $ on the transmission, for the same $k_0$ range and strength as in Fig.\ 
\ref{dia1}.}
\label{dia2}
\end{center}
\end{figure}

\begin{figure}
\begin{center}
\includegraphics[width=12cm]{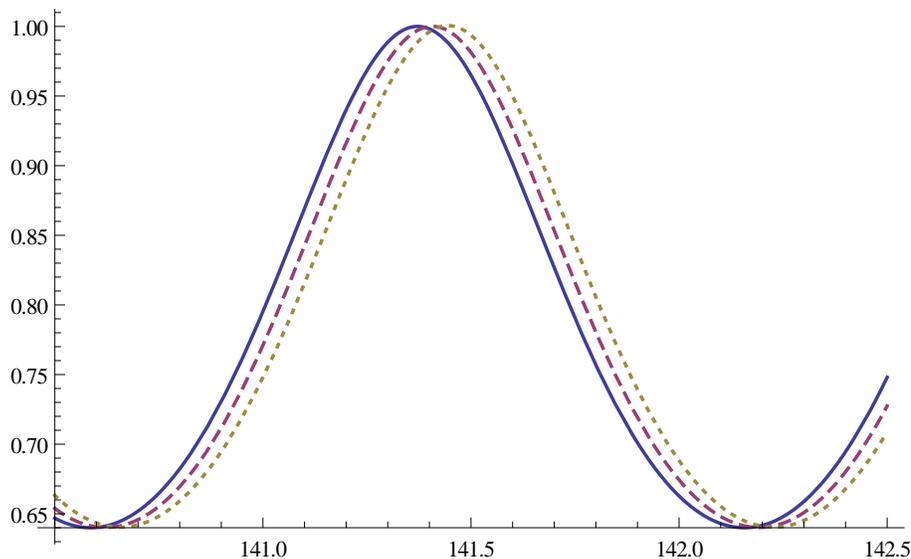}
\caption{ Same as in Fig.  \ref{dia2} but for  $\sigma k_0/\sqrt{\theta}=10 k_0L_0$.}
\label{dia3}
\end{center}
\end{figure}

The derived  expressions above are too complicated to provide a clear view of the effect of the noise on the transmission. In the present section we show a few  numerical computations of the current density corrections (\ref{cur1}-\ref{cur5}) for the loop graph. 
Note, that in order to avoid spurious effects resulting from the integration with respect to $\gamma_2$ from $-\infty$ to $\infty$ with the measure given in Eq.\ (\ref{measure}) as done in Eq.\ (\ref{aver}), we base the numerical results shown in the following on expressions obtained by integrations with respect to $\gamma_2$ from $-L_1$ to $L_2$  and normalize it by division by
\begin{equation}
\frac{\sqrt{2}\sigma}{\sqrt{\theta}}{\rm Erf}(\frac{\sqrt{\theta}L_2}{\sqrt{2}\sigma},-\frac{\sqrt{\theta}L_1}{\sqrt{2}\sigma})\equiv\frac{2}{\sqrt{\pi}}\int_{-L_1}^{L_2} d\gamma_2\,{\rm e}^{-\frac{\theta\gamma_2^2}{2\sigma^2}}.
\end{equation}
In our numerical illustration of Eqs.\ (\ref{cur1}-\ref{cur5}), we first consider $L={\hat L}=2L_0$ with $L_0$ a quantity with the dimension of a length, in this case 
the resonances are at $k=(n\pi/2-i\ln3/2)/L_0$ with $n\in\mathbb{N}$. We consider first the resonance at $k=(45\pi-i\ln3/2)/L_0$ and study its $k$-dependence for 
different values of $\alpha$ and $\sigma^2/\theta$. We consider in Fig. \ref{dia1} $\sigma^2k_0^2/\theta=0$ and $m\alpha/(\hbar^2k_0)=20/(k_0L_0)$ and 
$m\alpha/(\hbar^2k_0)=40/(k_0L_0)$ in the different curves. We observe that the resonance is shifted in $k$ by the perturbation with the shift increasing with increasing
strength of the perturbation and that it looses its symmetry with respect to its maximum. Considering small nonzero values for $\sigma^2/\theta$, 
$\sigma k_0/\sqrt{\theta}=0.005k_0L_0$ in Fig.\ \ref{dia2}, we obtain when considering the same $\alpha$-values as in Fig.\ \ref{dia1} again a shift and an asymmetry 
of the resonance that are however reduced compared to Fig.\ \ref{dia1}. For larger values of $\sigma^2/\theta$,  $\sigma k_0 /\sqrt{\theta}=10k_0L_0$ in 
Fig.\ \ref{dia3}, the size of the shift of the resonance is even further reduced and the asymmetry with respect to the maximum is completely gone.  

\begin{figure}
\begin{center}
\includegraphics[width=12cm]{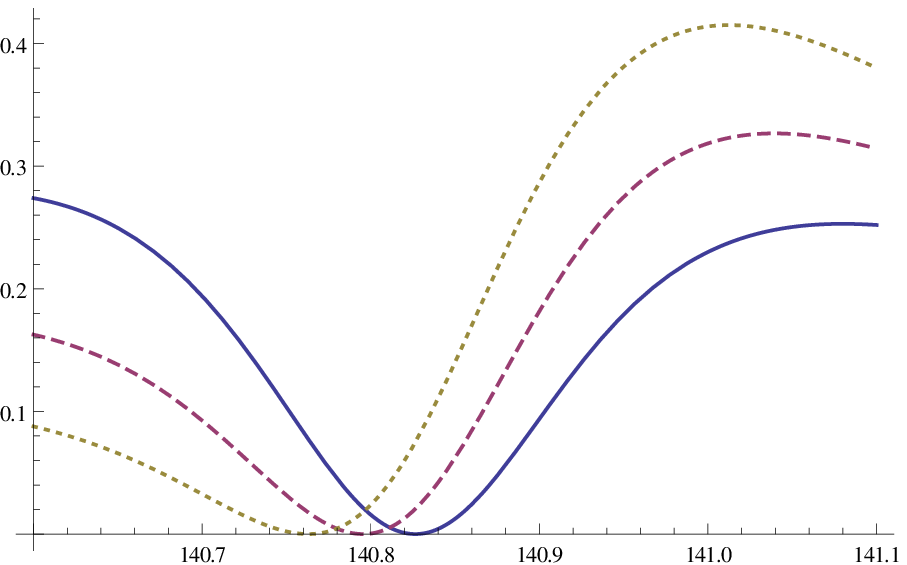}
\caption {The $k$-dependence of the transmission through  a loop due to a {\it static} perturbation, for $kL_0\in[140.6,141.1]$ -the vicinity of a narrow (topological)
resonance.  Three cases are shown: unperturbed transmission $m\alpha/(\hbar^2k_0)=0$  (blue full line), $m\alpha/(\hbar^2k_0)=14/(k_0L_0)$ (violet dashed line), 
$m\alpha/(\hbar^2k_0)=26/(k_0L_0)$ (brown dotted line).}
\label{dia4}
\end{center}
\end{figure}

\begin{figure}
\begin{center}
\includegraphics[width=12cm]{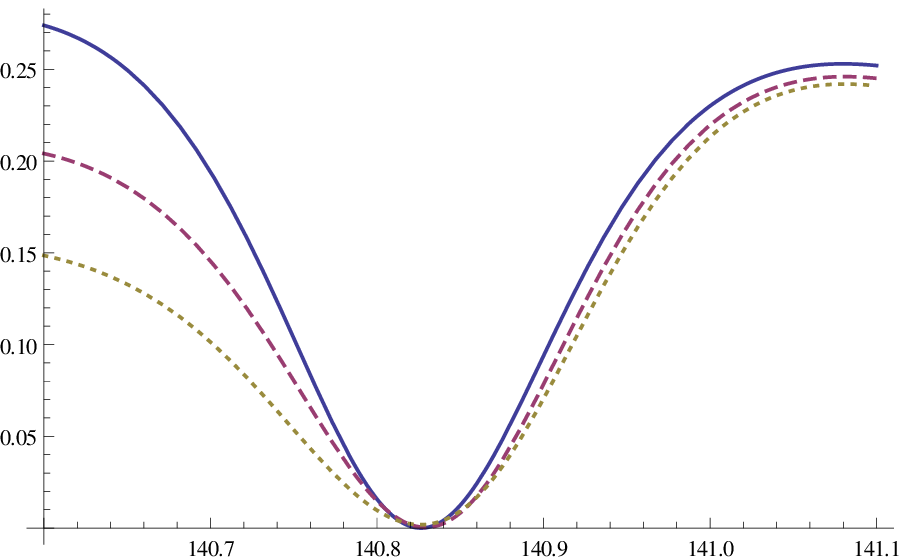}
\caption{The effect of noise characterized by  $\sigma k_0/\sqrt{\theta}=0.005 k_0L_0 $ on the transmission, for the same $k_0$ range and strength as in Fig.\ \ref{dia4}.}
\label{dia5}
\end{center}
\end{figure}

\begin{figure}
\begin{center}
\includegraphics[width=12cm]{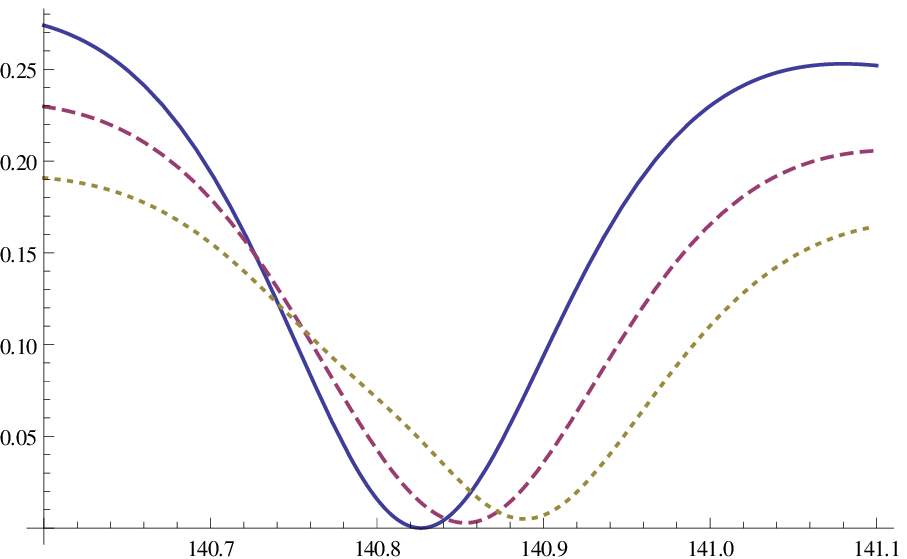}
\caption{ Same as in Fig.  \ref{dia5} but for  $\sigma k_0/\sqrt{\theta}=10 k_0L_0$.}
\label{dia6}
\end{center}
\end{figure}

Next we analyze the same graph for $L=(2+\pi/30)L_0$ and ${\hat L}=2L_0$. In the case of a small length difference between $L$ and ${\hat L}$ the bound states existing 
on the ring of the graph in Fig.\ \ref{fig1} for $L={\hat L}$ get weakly coupled to the leads \cite{Waltner}. This leads to resonances with much smaller width, so 
called topological resonances \cite{Schan}. For these bond lengths the denominators in Eq.\ (\ref{eq19}) possess zeros that come in pairs with both elements in the 
pair having approximately the same real parts \cite{Waltner}.  We consider the resonances situated at $k\approx(140.8 -0.14\,i)/L_0$ and $k\approx(140.8-0.40\,i)/L_0$ 
and describe again its shape in dependence of $\alpha$ and $\sigma^2$. We plot first its $k$-dependence for $\sigma^2/\theta=0$ and $m\alpha/(\hbar^2k_0)=0$, $m\alpha/
(\hbar^2k_0)=14/(k_0L_0)$ and $m\alpha/(\hbar^2k_0)=26/(k_0L_0)$ in Fig.\ \ref{dia4}. For $\sigma^2/\theta=0$ 
the effect of the perturbation is similar as for the broad resonance; also here a shift and the lost symmetry with respect to its maximum are observed that both 
decrease with increasing $\sigma^2/\theta$, see Figs.\ \ref{dia5}, \ref{dia6}. In this regime we obtain here additionally compared to the broad resonance a change of the shape of the resonance, i.e.\ a reduction of the overall height of the resonance and a broadening.

\begin{figure}
\begin{center}
\includegraphics[width=12cm]{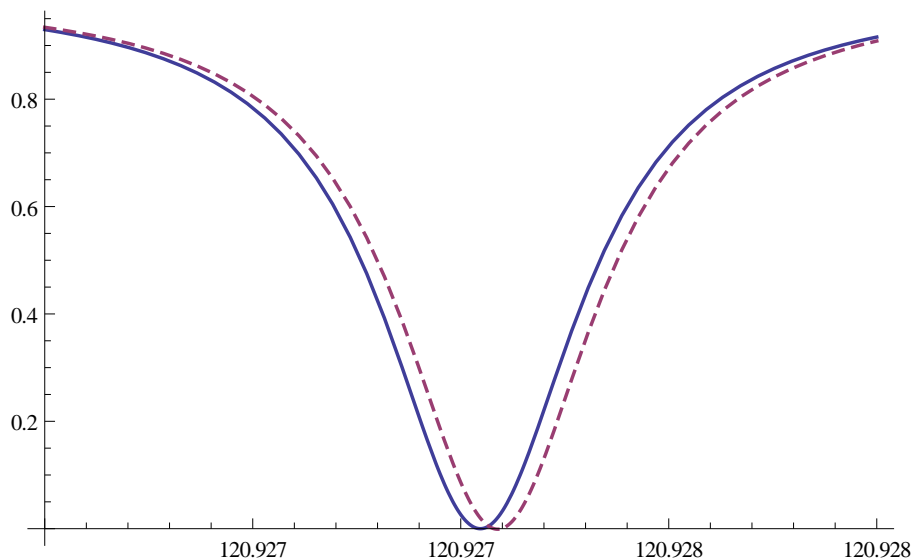}
\caption{The $k$-dependence of the transmission in the vicinity of a very narrow topological resonance in the regime $kL_0\in[120.926,120.928]$. The unperturbed case 
$m\alpha/(\hbar^2k_0)=0$ (blue full line) is compared with the transmission through a noisy graph with $m\alpha/(\hbar^2k_0)=0.02/(k_0L_0)$ and $\sigma k_0/
\sqrt{\theta}=10k_0L_0$ (violet dashed line).}
\label{dia10}
\end{center}
\end{figure}

For a sharper topological resonance smaller values of $\alpha$ were necessary to obtain an effect (shift, asymmetry) of a comparable size as for for the broad resonance. This is especially pronounced
for a very sharp topological resonance shown without perturbation as blue full line in Fig.\ \ref{dia10}: The violet dashed curve is obtained for $m\alpha/(\hbar^2k_0)=0.02/(k_0L_0)$ and $\sigma k_0/\sqrt{\theta}=10 k_0L_0$. Unfortunately it is not possible to increase $\alpha$ further in this case as the perturbation theory applied breaks down in this regime which manifests by the fact that negative values are obtained for the transmission. Thus it is not possible to observe an expected broadening of the resonance as in Fig.\ \ref{dia6}.
\begin{figure}
\begin{center}
\includegraphics[width=12cm]{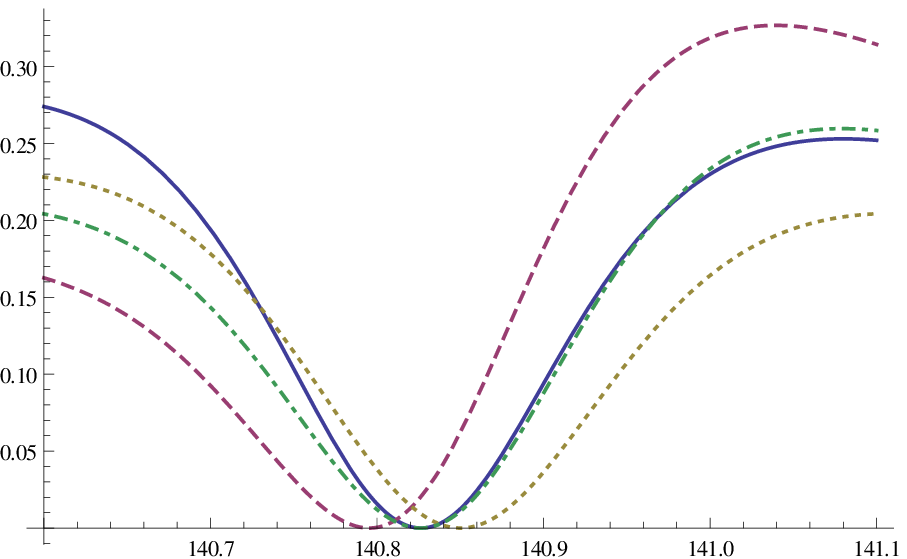}
\caption{The effect of a {\it static} delta potential for $m\alpha/\hbar^2k_0=14/(k_0L_0)$ on the resonance from Fig.\ \ref{dia4}, for $L_1=1L_0$ (violet dashed line), 
$L_1=0.95L_0$ (brown dotted line), $L_1=0.87$ (green dashed dotted line) and $L$, $\hat{L}$ fixed. The unperturbed resonance is shown (blue full line).}
\label{dia9}
\end{center}
\end{figure}

We obtain a better intuitive understanding of the behavior observed by noting that a {\it static} perturbation leads essentially to asymmetric shifts of the resonance positions 
with the size of the shift depending on the position where the delta potential is placed on bond "2", see Fig.\ \ref{dia9}, where the transmission spectrum is shown for different values of 
of $L_1$ in the different curves with $L$ and $\hat{L}$ fixed for the same resonance as in Fig.\ \ref{dia4}-\ref{dia6}. The noise average now averages over all 
these possible configurations. This leads to a reduction in height of the resonances near transmission value $|T(k_0)|^2=0$ as an average over the transmission curves 
shown in Fig.\ \ref{dia9} leads at that place to an enhanced transmission.

\section{Details on the derivation of the results given above}\label{sect5}
In this section we explain mainly technical details leading to the results explained in sections \ref{sections4} and \ref{sect3}.
\subsection{Current through a delta potential at a randomly fluctuating position}
We calculate $\left\langle\Psi^{(1)}(x,t)\right\rangle$ and $\left\langle\Psi^{(2)}(x,t)\right\rangle$ for the setting of a single delta potential fluctuating randomly in position around $x=0$ as introduced in Eq.\ (\ref{eq3000}).

Starting with $\left\langle\Psi^{(1)}(x,t)\right\rangle$ we obtain for the noise average of the $\gamma(t')$ from Eq.\ (\ref{aver})
\begin{equation}
\label{eq1004}
\left\langle{\rm e}^{i(k_0-k')\gamma(t')}\right\rangle={\rm e}^{-\frac{\sigma^2}{2\theta}(k_0-k')^2}.
\end{equation}
Inserting this averaged noise contribution into Eq.\ (\ref{eq6aa}) and performing the $t'$-integral we get
\begin{eqnarray}
\label{eq1005}
&\fl\left\langle\Psi^{(1)}(x,t)\right\rangle=\lim_{\epsilon\rightarrow 0}\frac{\alpha}{2\pi i\hbar}\int_{-\infty}^\infty dk'\int_{0}^tdt'{\rm e}^{-i\hbar k'^2t/(2m)+ik'x}
{\rm e}^{i\left(\hbar k'^2/(2m)-\hbar k_0^2/(2m)-i\epsilon\right)t'}{\rm e}^{-\frac{\sigma^2}{2\theta}(k_0-k')^2}\nonumber\\&\fl=-
\lim_{\epsilon\rightarrow 0}\frac{\alpha}{2\pi \hbar}\int_{-\infty}^\infty dk' \frac{{\rm e}^{-i\hbar k_0^2t/(2m)+ik'x}}{\frac{\hbar}{2m}(k'^2-k_0^2)-i\epsilon}
\left[{\rm e}^{-\frac{\sigma^2}{2\theta}(k_0-k')^2}-{\rm e}^{-i\hbar (k'^2-k_0^2)t/2m}\right].
\end{eqnarray}
The remaining $k'$-integral we will perform by residual integration, its result depends on the considered regime: for small times $t$ and large $x$-values the 
perturbation has not yet any effect at considered position. In this case we need to close the integration contour in the upper half plane, the two contributions in the 
curved bracket cancel. Increasing $t$, we need to integrate the first term in the curved bracket in the same way yielding no contribution and choose for the second term a contour situated such 
that the product of real and imaginary part of $k'$ is negative. This regime will always be considered in the following.
Finally, this yields
\begin{equation}\label{firsto}
\left\langle\Psi^{(1)}(x,t)\right\rangle=-\frac{i\alpha m}{\hbar^2k_0}{\rm e}^{-i\hbar k_0^2t/(2m)+ik_0x}
\end{equation}
and by this a purely imaginary contribution to $\Psi^{(0)*}(x,t)\partial_x\left\langle\Psi^{(1)}(x,t)\right\rangle$ and no contribution to $\Delta j^{(1)}(x,t)$ given 
in Eq.\ (\ref{eq5000}). The quantity $\left\langle\Psi^{(1)*}(x,t)\partial_x\Psi^{(1)}(x,t)\right\rangle$ can be obtained by noting that our averages can be considered 
as two independent Gaussian averages yielding
\begin{equation}
\label{split}
\left\langle\Psi^{(1)*}(x,t)\partial_x\Psi^{(1)}(x,t)\right\rangle=\left\langle\Psi^{(1)*}(x,t)\right\rangle\partial_x\left\langle
\Psi^{(1)}(x,t)\right\rangle
\end{equation}

The last quantity to be analyzed is $\left\langle\Psi_2(x,t)\right\rangle$. Here we get from Eq.\ (\ref{eq9}) when performing the noise average by means of Eq.\ (\ref{aver})
\begin{eqnarray}
\label{eq1006}
&\fl\left\langle\Psi^{(2)}(x,t)\right\rangle=-\lim_{\epsilon\rightarrow 0}\frac{\alpha^2}{4\pi^2\hbar^2}\int_{-\infty}^\infty dk_1dk_2\int_{0}^tdt_2\int_{0}^{t_2}
dt_1{\rm e}^{i[\hbar k_1^2/(2m)-i\epsilon](t_1-t_2)}{\rm e}^{ik_2x}\nonumber\\&\fl\times{\rm e}^{i[\hbar k_2^2/(2m)-i\epsilon](t_2-t)-i\hbar k_0t_1/(2m)}
{\rm e}^{-\sigma^2(k_1-k_2)^2/(2\theta)-\sigma^2(k_0-k_1)^2/(2\theta)}.
\end{eqnarray}
After performing the time integrations we get
\begin{eqnarray}
\label{eq1009}
&\fl\left\langle\Psi^{(2)}(x,t)\right\rangle=\lim_{\epsilon\rightarrow 0}\frac{\alpha^2}{4\pi\hbar^2}\int_{-\infty}^\infty dk_1dk_2\frac{{\rm e}^{ik_2x}
{\rm e}^{-\sigma^2(k_1-k_2)^2/(2\theta)-\sigma^2(k_0-k_1)^2/(2\theta)}}{\frac{\hbar}{2m}(k_1^2-k_0^2)-i\epsilon}\nonumber\\&\fl\times
\left\{\frac{\left[{\rm e}^{-i\hbar k_0^2t/(2m)}-{\rm e}^{-i\hbar k_2^2t/(2m)}\right]}{\left[\frac{\hbar}{2m}(k_2^2-k_0^2)-i\epsilon\right]}
-\frac{\left[{\rm e}^{-i\hbar k_1^2t/(2m)}-{\rm e}^{-i\hbar k_2^2t/(2m)}\right]}{\frac{\hbar}{2m}(k_2^2-k_1^2)}\right\}.
\end{eqnarray}
Again only the first term in the curved brackets is relevant, here we close the integration contour of the $k_2$-integration in the upper half plane. For the $k_1$-integration we note that only the real part of ${\rm e}^{-ik_0x}{\rm e}^{i\hbar k_0^2t/(2m)}\left\langle\Psi^{(2)}(x,t)\right\rangle$ is needed for computing the current contribution implying that in
\begin{equation}
\lim_{\epsilon\rightarrow 0}\frac{1}{\frac{\hbar}{2m}(k_1^2-k_0^2)-i\epsilon}=\frac{2m}{\hbar}\left[i\pi\delta\left(k_1^2-k_0^2\right)+\mathcal{P}\left(
\frac{1}{k_1^2-k_0^2}\right)\right]
\end{equation}
only the term containing the delta function will be relevant. We obtain
\begin{equation}
\label{eq1010}
\Re\left[{\rm e}^{i\hbar k_0^2t/(2m)}{\rm e}^{-ik_0x}\left\langle\Psi^{(2)}(x,t)\right\rangle\right]=-\frac{\alpha^2m^2}{2\hbar^4k_0^2}\left(1+{\rm e}^{-2\sigma^2k_0^2/\theta}\right).
\end{equation}
We now exchange the unperturbed system, a freely propagating plane wave above, against a quantum graph.

\subsection{Current through a noisy graph}
Here give details on the calculation of the effect of a fluctuating delta potential located on a quantum graph.  
Again we consider $\left\langle\Psi^{(1)}(x,t)\right\rangle$, $\left\langle\Psi^{(1)*}(x,t)\partial_x\Psi^{(1)}(x,t)\right\rangle$ and $\left\langle\Psi^{(2)}(x,t)\right\rangle$.

Starting from Eq.\ (\ref{eq1030}) we get for $\left\langle\Psi^{(1)}(x,t)\right\rangle$:
\begin{eqnarray}
\label{eq1032}
&\fl\left\langle\Psi^{(1)}(x,t)\right\rangle
=\frac{\alpha}{2\pi i\hbar}\lim_{\epsilon\rightarrow 0}\int_{0}^\infty dk'\int_{0}^tdt' {\rm e}^{i\left(\hbar k'^2/(2m)-\hbar k_0^2/(2m)-i\epsilon\right)t'}{\rm e}^{-i\hbar k'^2t/(2m)+ik'x}C(k')
\nonumber\\&\fl\times\left\{{\rm e}^{-\frac{\sigma^2}{2\theta}(k_0-k')^2}\left[A(k_0)A^*(k'){\rm e}^{i(k_0-k')L_1}+B(k_0)B^*(k'){\rm e}^{-i(k_0-k')L_1}\right]\right.
\nonumber\\&\fl\left.+{\rm e}^{-\frac{\sigma^2}{2\theta}(k_0+k')^2}\left[A(k_0)B^*(k'){\rm e}^{i(k_0+k')L_1}+A^*(k')B(k_0){\rm e}^{-i(k_0+k')L_1}\right]\right\}\nonumber\\&\fl+\frac{\alpha}{2\pi i\hbar}
\lim_{\epsilon\rightarrow 0}\int_{0}^\infty dk'\int_{0}^tdt'{\rm e}^{-i\hbar k'^2t/(2m)+ik'x} {\rm e}^{i\left(\hbar k'^2/(2m)-\hbar k_0^2/(2m)-i\epsilon\right)t'}({\rm e}^{-ik'x}+R(k'){\rm e}^{ik'x})
\nonumber\\&\fl\times\left\{{\rm e}^{-\frac{\sigma^2}{2\theta}(k_0-k')^2}\left[A(k_0)A^*(k'){\rm e}^{ik_0L_1-ik'L_2}+B(k_0)B^*(k')
{\rm e}^{-ik_0L_1+ik'L_2}\right]\right.\nonumber\\&\fl\left.+{\rm e}^{-\frac{\sigma^2}{2\theta}(k_0+k')^2}\left[A(k_0)B^*(k'){\rm e}^{ik_0L_1+ik'L_2}+A^*(k')B(k_0)
{\rm e}^{-ik_0L_1-ik'L_2}\right]\right\}\nonumber\\&\fl=-\lim_{\epsilon\rightarrow 0}
\frac{\alpha}{2\pi\hbar}\int_{0}^\infty \!\!\!dk'{\rm e}^{-i\hbar k_0^2t/(2m)+ik'x} \left\{\left[A(k_0)A^*(k'){\rm e}^{i(k_0-k')L_1}+B(k_0)B^*(k'){\rm e}^{-i(k_0-k')L_1}\right]\right.
\nonumber\\&\fl\left.\times\frac{{\rm e}^{-\frac{\sigma^2}{2\theta}(k_0-k')^2}}{\frac{\hbar}{2m}(k'^2-k_0^2)
-i\epsilon}+\left[A(k_0)B^*(k'){\rm e}^{i(k_0+k')L_1}+A^*(k')B(k_0){\rm e}^{-i(k_0+k')L_1}\right]\right.\!\!\nonumber\\&\fl \left.\times\frac{{\rm e}^{-\frac{\sigma^2}{2\theta}(k_0+k')^2}}{\frac{\hbar}{2m}(k'^2-k_0^2)
-i\epsilon}\right\}C(k')-\lim_{\epsilon\rightarrow 0}
\frac{\alpha}{2\pi\hbar}\int_{0}^\infty dk'{\rm e}^{-i\hbar k_0^2t/(2m)}({\rm e}^{-ik'x}+R(k'){\rm e}^{ik'x})\nonumber\\&\fl\times\left\{\frac{{\rm e}^{-\frac{\sigma^2}{2\theta}(k_0+k')^2}}{\frac{\hbar}{2m}(k'^2-k_0^2)
-i\epsilon}\left[A(k_0)A^*(k'){\rm e}^{ik_0L_1-ik'L_2}+B(k_0)B^*(k'){\rm e}^{-ik_0L_1+ik'L_2}\right]
 \right.\nonumber\\&\fl\left.+\frac{{\rm e}^{-\frac{\sigma^2}{2\theta}(k_0-k')^2}}{\frac{\hbar}{2m}(k'^2-k_0^2)
-i\epsilon}\left[A(k_0)B^*(k'){\rm e}^{ik_0L_1+ik'L_2}+A^*(k')B(k_0){\rm e}^{-ik_0L_1-ik'L_2}\right]\right\}
\nonumber\\&\fl
=-\frac{i\alpha m}{\hbar^2k_0}{\rm e}^{-i\hbar k_0^2t/(2m)}{\rm e}^{ik_0x}\left\{C(k_0)\left[|A(k_0)|^2+|B(k_0)|^2+{\rm e}^{-2\sigma^2k_0^2/\theta}\right.\right.
\nonumber\\&\fl\left.\left.\times\left(A(k_0)B^*(k_0){\rm e}^{2ik_0L_1}+A^*(k_0)B(k_0){\rm e}^{-2ik_0L_1}\right)\right]+R(k_0)\left[A^*(k_0)B(k_0){\rm e}^{-ik_0(L_2+L_1)}\right.\right.\nonumber\\&\fl\left.
\left.+B^*(k_0)A(k_0){\rm e}^{ik_0(L_2+L_1)}+{\rm e}^{-2\sigma^2k_0^2/\theta}\left(|A(k_0)|^2{\rm e}^{-ik_0(L_2-L_1)}+|B(k_0)|^2{\rm e}^{ik_0(L_2-L_1)}\right)\right]\right\},
\end{eqnarray}
where we omitted in the second step the contributions from the lower limit of the time-integrations as they cancel in the same way as in the previous subsection.
Contrary to the last subsection we get here from the term proportional to $R(k_0)$ a contribution linear in $\alpha$ to the current as here ${\rm e}^{-ik_0x}{\rm e}^{i\hbar k_0^2t/(2m)}C^*(k_0)\left\langle\Psi_1(x,t)\right\rangle$ is not real. 

The quantity $\left\langle\Psi^{(1)*}(x,t)\partial_x\Psi^{(1)}(x,t)\right\rangle$ we obtain again by using relation (\ref{split}).

Finally we evaluate the contributions originating from $\left\langle\Psi^{(2)}(x,t)\right\rangle$:
We start with the term in (\ref{eq1031}) containing the factor $A(k_0)A^*(k_1)A(k_1)A^*(k_2)$ that we denote by $\Psi^{(2)}_a(x,t)$. We label the different contributions with roman letters in alphabetical order. Performing the $t_i$-integrals as in the last subsection, we  obtain
\begin{eqnarray}
\label{eq1035}
&\fl\left\langle\Psi^{(2)}_a(x,t)\right\rangle=\lim_{\epsilon\rightarrow 0}\frac{\alpha^2}{4\pi^2\hbar^2}\int_{-\infty}^\infty dk_1\int_0^\infty dk_2
\frac{{\rm e}^{-i\hbar k_0^2t/(2m)}A(k_0)|A(k_1)|^2}{\left[\frac{\hbar}{2m}(k_1^2-k_0^2)-i\epsilon\right]}\frac{A^*(k_2){\rm e}^{i(k_0-k_2)L_1}}{\left[\frac{\hbar}
{2m}(k_2^2-k_0^2)-i\epsilon\right]}\nonumber\\&\fl\times\left[C(k_2){\rm e}^{ik_2x}{\rm e}^{-\sigma^2\left[(k_1-k_0)^2+(k_1-k_2)^2\right]/(2\theta)}+(R(k_2){\rm e}^{ik_2 x}+{\rm e}^{-ik_2 x})\right.\nonumber\\&\fl\left.\times
{\rm e}^{-ik_2(L_2-L_1)}{\rm e}^{-\sigma^2\left[(k_1-k_0)^2-(k_1+k_2)^2\right]/(2\theta)}\right],
\end{eqnarray}
where we again omitted the contributions from the other limits of the time integral.
The $k_2$-integral is performed in the upper complex plane with the contribution resulting from the pole at $k_2=k_0$ yielding
\begin{eqnarray}
\label{eq1036}
&&\left\langle\Psi^{(2)}_a(x,t)\right\rangle=\frac{\alpha^2mi}{2\pi\hbar^3k_0}\int_{-\infty}^\infty dk_1\frac{{\rm e}^{-i\hbar k_0^2t/(2m)}{\rm e}^{ik_0x}|A(k_0)|^2|
A(k_1)|^2}{\frac{\hbar}{2m}(k_1^2-k_0^2)-i\epsilon}\\&&\times\left[C(k_0){\rm e}^{-\sigma^2(k_1-k_0)^2/\theta}+R(k_0){\rm e}^{-ik_0(L_2-L_1)}{\rm e}^{-\sigma^2\left[(k_1-k_0)^2+(k_1+k_0)^2\right]/(2\theta)}\right]\nonumber
\end{eqnarray}
When performing the $k_1$-integral the contribution relevant for the current density resulting from the term proportional to $C(k_0)$ is obtained in the same way as explained after Eq.\ (\ref{eq1009}). For the term proportional to $R(k_0)$ we rewrite the Gaussians in the following way
\begin{eqnarray}
\label{aux}
{\rm e}^{-\sigma^2\left[(k_1-k_0)^2+(k_1+k_0)^2\right]/(2\theta)}=\int_{-\infty}^\infty dxdy\, {\rm e}^{-x^2-y^2+\sqrt{2}\sigma i\left[(k_1-k_0)x+(k_1+k_0)y\right]/\sqrt{\theta}}.
\end{eqnarray}
Inserting this in Eq.\ (\ref{eq1036}), we split the $x$- and $y$-integral into the regions $x+y>0$ and $x+y<0$. In the first region we close the contour for the 
$k_1$-integral in the upper and in the second in the lower half plane, respectively. In the upper half plane we have the pole at $k_1=k_0+i\frac{\epsilon'}{k_0}$ with 
$\epsilon'$ a small positive constant and 
the pole of $A(k_1)$ at $k_1=\tilde{k}_1$ that we assume to be situated in the upper right complex plane. The latter pole implies that there is also a pole of $A(k)$ 
at $-\tilde{k}_1^*$ as also the complex conjugate of the denominator of $A(k)$ vanishes at $\tilde{k}_1$. In the lower half plane we have a pole at $k_1=-k_0-i\frac{\epsilon'}{k_0}$. 
The function $A^*(k_1)$ has poles at $\tilde{k}_1^*$ and $-\tilde{k}_1$. We make use here of the notation $A(k)=A_o(k)/\mathcal{D}(k)$ introduced after Eq.\ (\ref{cur5}). 
Performing the $x$- and $y$-integrals afterwards we get
\begin{eqnarray}
\label{eq1037}
&\fl\Re\left[C^*(k_0){\rm e}^{-ik_0x}{\rm e}^{i\hbar k_0^2t/(2m)}\left\langle \Psi^{(2)}_a(x,t)\right\rangle\right]=-\frac{\alpha^2m^2}{2\hbar^4k_0^2}|A(k_0)|^2
\Re\left\{|C(k_0)|^2|A(k_0)|^2\right.\\&\fl\left.\times\left(1+{\rm e}^{-4\sigma^2k_0^2/\theta}\right)+2C^*(k_0)R(k_0){\rm e}^{-ik_0(L_2-L_1)}|A(k_0)|^2
{\rm e}^{-2\sigma^2k_0^2/\theta}\left(1+{\rm Erf}(i\sigma k_0/\sqrt{\theta})\right)+\nonumber\right.
\\&\fl\left.
8C^*(k_0)R(k_0){\rm e}^{-ik_0(L_2-L_1)}\Im\left[\frac{k_0}{\tilde{k}_1^2-k_0^2} A^*(\tilde{k}_1)\frac{A_o(\tilde{k}_1)}{\mathcal{D}_u'(\tilde{k}_1)}
{\rm e}^{-\sigma^2(\tilde{k}_1^2+k_0^2)/\theta}\left(1+{\rm Erf}(i\sigma\tilde{k}_1/\sqrt{\theta})\right)\right]\right\}\nonumber
\end{eqnarray}
with the error function defined as ${\rm Erf}(x)\equiv\frac{2}{\sqrt{\pi}}\int_0^xdy{\rm e}^{-y^2}$.
All other contributions resulting from Eq.\ (\ref{eq1031}) can be evaluated in a similar manner, the changes are that at least some of the $A(k)$ are replaced by $B(k)$ and the $\sigma$-dependent Gaussians are altered. We will only give for each contribution the form of the $\sigma$-dependent Gaussian and the final result.
Next we consider the contribution in Eq.\ (\ref{eq1031}) containing the factor $B(k_0)B^*(k_1)B(k_1)B^*(k_2)$. 
The calculation is completely analogous to the one above as there is no change in the $\sigma$-dependent Gaussians and we obtain the result by replacing $A(k)$ by $B(k)$ and complex conjugating ${\rm e}^{-ik_0(L_2-L_1)}$ in 
the last equation. 
The next term results from the summand in (\ref{eq1031}) containing $A(k_0)A^*(k_1)A(k_1)B^*(k_2)$. Here we have in contrast to Eq.\ (\ref{eq1036}) the $\sigma$-
dependent Gaussians ${\rm e}^{-\sigma^2\left[(k_1-k_0)^2+(k_1+k_0)^2\right]/(2\theta)}$ in the term proportional to $C(k_0)$ and 
${\rm e}^{-\sigma^2(k_1-k_0)^2/\theta}$ in the term proportional to $R(k_0)$. For the last term we again introduce an auxiliary integral as explained in 
Eq.\ (\ref{aux}) and finally get
\begin{eqnarray}
\label{eq1038}
&\fl\Re\left[C^*(k_0){\rm e}^{-ik_0x}{\rm e}^{i\hbar k_0^2t/(2m)}\left\langle\Psi^{(2)}_b(x,t)\right\rangle\right]=-\frac{\alpha^2m^2}{2\hbar^4k_0^2}\Re \left\{A(k_0)B^*(k_0)
{\rm e}^{2ik_0L_1}\left\{2|C(k_0)|^2
\right.\right.\nonumber\\&\fl\left.\left.\times|A(k_0)|^2{\rm e}^{-2\sigma^2k_0^2/\theta}+C^*(k_0)R(k_0){\rm e}^{ik_0(L_2-L_1)}|A(k_0)|^2\left[
{\rm e}^{-4\sigma^2k_0^2/\theta}\left(1+{\rm Erf}(2i\sigma k_0/\sqrt{\theta})\right)\right.\right.\right.\nonumber\\&\fl\left.\left.\left.+1\right]+4C^*(k_0)R(k_0){\rm e}^{ik_0(L_2-L_1)}\Im\left[\frac{k_0}{(\tilde{k}_1^2-k_0^2)}A^*(\tilde{k}_1)\frac{A_o(\tilde{k}_1)}{\mathcal{D}_u'(\tilde{k}_1)}
\left[\left(1+{\rm Erf}(i\sigma(\tilde{k}_1-k_0)/\sqrt{\theta})\right)
\right.\right.\right.\right.\nonumber\\&\fl\left.\left.\left.\left.\times{\rm e}^{-\sigma^2(\tilde{k}_1-k_0)^2/\theta}+{\rm e}^{-\sigma^2(\tilde{k}_1+k_0)^2/\theta}\left(1+ {\rm Erf}(i\sigma(
\tilde{k}_1+k_0)/\sqrt{\theta})\right)\right]\right]\right\}\right\}.
\end{eqnarray}
The same $\sigma$-dependent Gaussians are obtained for the summand in Eq.\ (\ref{eq1031}) containing $B(k_0)B^*(k_1)B(k_1)A^*(k_2)$, this contribution is obtained by exchanging $A(k)$ and $B(k)$ and complex conjugating ${\rm e}^{ik_0(L_2-L_1)}$, ${\rm e}^{2ik_0L_1}$ in the last equation.
Considering next the term proportional to $B(k_0)A^*(k_1)A(k_1)A^*(k_2)$ we have the same $\sigma$-dependent Gaussian as in the last contribution in the term proportional to $C(k_0)$ and the term ${\rm e}^{-\sigma^2(k_0+k_1)^2/\theta}$ in the term proportional to $R(k_0)$ in Eq.\ (\ref{eq1037}) and get
\begin{eqnarray}
\label{eq10037}
&\fl\Re\left[C^*(k_0){\rm e}^{-ik_0x}{\rm e}^{i\hbar k_0^2t/(2m)}\left\langle\Psi^{(2)}_c(x,t)\right\rangle\right]=-\frac{\alpha^2m^2}{2\hbar^4k_0^2}\Re\left\{A^*(k_0)B(k_0){\rm e}^{-2ik_0L_1}
\left\{2|C(k_0)|^2\right.\right.\nonumber\\&\fl\left.\left.\times|A(k_0)|^2
{\rm e}^{-2\sigma^2k_0^2/\theta}+C^*(k_0)R(k_0)|A(k_0)|^2{\rm e}^{-ik_0(L_2-L_1)}\left[{\rm e}^
{-4\sigma^2k_0^2/\theta}\left(1+{\rm Erf}(2i\sigma k_0/\sqrt{\theta})\right)\right.\right.\right.\nonumber\\&\fl\left.\left.\left.+1\right]+4C^*(k_0)R(k_0){\rm e}^{-ik_0(L_2-L_1)}\Im\left[\frac{k_0}{(\tilde{k}_1^2-k_0^2)}A^*(\tilde{k}_1)\frac{A_o(\tilde{k}_1)}{\mathcal{D}_u'(\tilde{k}_1)}
\left[\left(1+{\rm Erf}(i\sigma(\tilde{k}_1+k_0))\right)\right.\right.\right.\right.
\nonumber\\&\fl\left.\left.\left.\left.\times{\rm e}^{-\sigma^2(\tilde{k}_1+k_0)^2/\theta}+{\rm e}^{-\sigma^2(\tilde{k}_1-k_0)^2/\theta}\left(1+{\rm Erf}(i\sigma
(\tilde{k}_1-k_0))\right)\right]\right]\right\}\right\}.
\end{eqnarray}
The contribution containing the factor $A(k_0)B^*(k_1)B(k_1)B^*(k_2)$ is obtained from the last one by exchanging $A(k)$ and $B(k)$ and complex conjugating ${\rm e}^{-ik_0(L_2-L_1)}$, ${\rm e}^{-2ik_0L_1}$.
The next term results from the summand in (\ref{eq1031}) containing $A(k_0)B^*(k_1)B(k_1)A^*(k_2)$, here the $\sigma$-dependent Gaussians are exchanged compared to Eq.\ (\ref{eq10037}) and get
\begin{eqnarray}
\label{eq10389}
&\fl\Re\left[C^*(k_0){\rm e}^{i\hbar k_0^2t/(2m)}{\rm e}^{-ik_0x}\left\langle\Psi^{(2)}_d(x,t)\right\rangle\right]=-\frac{\alpha^2m^2}{2\hbar^4k_0^2}|A(k_0)|^2\left\{|C(k_0)|^2|B(k_0)|^2
\right.\nonumber
\\&\fl\left.\times
\left(1+{\rm e}^{-4\sigma^2k_0^2/\theta}\right)+2C^*(k_0)R(k_0)|B(k_0)|^2{\rm e}^{-ik_0(L_2-L_1)}{\rm e}^{-2\sigma^2k_0^2/\theta}\left(1+{\rm Erf}(i\sigma\tilde{k}_0)
\right)\right.\\&\fl
\left.+8C^*(k_0)R(k_0){\rm e}^{-ik_0(L_2-L_1)}
\Im\left[\frac{k_0}{(\tilde{k}_1^2-k_0^2)}B^*(\tilde{k}_1)\frac{B_o(\tilde{k}_1)}
{\mathcal{D}_u'(\tilde{k}_1)}{\rm e}^{-\sigma^2(\tilde{k}_1^2+k_0^2)/\theta}\left(1+{\rm Erf}(i\sigma\tilde{k}_1)\right)\right]\right\}\nonumber.
\end{eqnarray}
The contribution resulting from the term containing $B(k_0)A^*(k_1)A(k_1)B^*(k_2)$ is obtained from the last equation by exchanging $A(k)$ and $B(k)$ and complex conjugating ${\rm e}^{-ik_0(L_2-L_1)}$.

In the case of the summands $A(k_0)A(k_1)B^*(k_1)A^*(k_2)$ and $A(k_0)B(k_1)A^*(k_1)A^*(k_2)$ in Eq.\ (\ref{eq1031}) we obtain the $\sigma$-dependent Gaussians 
${\rm e}^{-\sigma^2\left[(k_1+k_0)^2+(k_1-k_0)^2\right]/(2\theta)}$ for the terms proportional to $C(k_0)$ and ${\rm e}^{-\sigma^2(k_1-k_0)^2/(2\theta)}$ for the ones 
proportional to $R(k_0)$
\begin{eqnarray}
&\fl\Re\left[C^*(k_0){\rm e}^{-ik_0x}{\rm e}^{i\hbar k_0^2/(2m)}\left\langle\Psi_e(x,t)\right\rangle\right]=-\frac{\alpha^2m^2}{2\hbar^4k_0^2}\Re\left\{C^*(k_0)|A(k_0)|^2
\left\{C(k_0)\left[A(k_0)B^*(k_0)\right.\right.\right.\nonumber\\&\fl\left.\left.\left.\times\left({\rm e}^{2ik_0L_2}+{\rm e}^{2ik_0L_1}\right)+A^*(k_0)B(k_0)\left(
{\rm e}^{-2ik_0L_2}+{\rm e}^{-2ik_0L_1}\right)\right]
{\rm e}^{-2\sigma^2k_0^2/\theta}+R(k_0){\rm e}^{-ik_0(L_2-L_1)}\right.\right.\nonumber\\&\fl\left.\left.\times\left[A(k_0)B^*(k_0){\rm e}^{2ik_0L_2}+A^*(k_0)B(k_0){\rm e}^{-2ik_0L_1}+{\rm e}^{-4\sigma^2k_0^2/\theta}\left(1+{\rm Erf}(2i\sigma k_0/\sqrt{\theta})\right)\right.\right.\right.
\nonumber\\&\fl\left.\left.\left.\times
\left(A^*(k_0)B(k_0){\rm e}^{-2ik_0L_2}+A(k_0)B^*(k_0){\rm e}^{2ik_0L_1}\right)
\right]+4R(k_0){\rm e}^{-ik_0(L_2-L_1)}\Im\left[\frac{k_0}{\tilde{k}_1^2-k_0^2}\right.\right.\right.\nonumber\\&\fl\left.\left.\left.\times\left(B^*(\tilde{k}_1)\frac{A_o(\tilde{k}_1)}
{\mathcal{D}_u'(\tilde{k}_1)}{\rm e}^{2i\tilde{k}_1L_2}+A^*(\tilde{k}_1)\frac{B_o(\tilde{k}_1)}{\mathcal{D}_u'(\tilde{k}_1)}{\rm e}^{-2i\tilde{k}_1L_1}
\right){\rm e}^{-\sigma^2(\tilde{k}_1-k_0)^2/\theta}\left(1+\right.\right.\right.\right.\nonumber\\&\fl\left.\left.\left.\left.{\rm Erf}(i\sigma(\tilde{k}_1-k_0)/\sqrt{\theta})\right)+{\rm e}^{-\sigma^2(\tilde{k}_1+k_0)^2/\theta}
\left(A^*(\tilde{k}_1)\frac{B_o(\tilde{k}_1)}{\mathcal{D}_u'(\tilde{k}_1)}{\rm e}^{-2i\tilde{k}_1L_2}+B^*(\tilde{k}_1)\frac{A_o(\tilde{k}_1)}{\mathcal{D}_u'(\tilde{k}_1)}{\rm e}^{2i\tilde{k}_1L_1}\right)
\right.\right.\right.\nonumber\\&\fl \left.\left.\left.\times\left(1+{\rm Erf}(i\sigma(\tilde{k}_1+k_0)/\sqrt{\theta})\right)\right]\right\}\right\}
\end{eqnarray}
The contribution resulting from $B(k_0)B(k_1)A^*(k_1)B^*(k_2)$ and $B(k_0)A(k_1)B^*(k_1)B^*(k_2)$ in Eq.\ (\ref{eq1031}) are obtained again from the last result by the exchange of $A(k)$ and $B(k)$ and complex conjugation of  the $\sigma$-independent exponentials. For the contributions in Eq.\ (\ref{eq1031}) $B(k_0)B(k_1)A^*(k_1)A^*(k_2)$ and $B(k_0)A(k_1)B^*(k_1)A^*(k_2)$ we obtain the $\sigma$-dependent Gaussians ${\rm e}^{-\sigma^2(k_1+k_0)^2/\theta}$ in the term proportional to $C(k_0)$ and ${\rm e}^{-\sigma^2\left[(k_1-k_0)^2+(k_1+k_0)^2\right]/\theta}$ in the term proportional to $R(k_0)$. We get for the contribution to the wave function
\begin{eqnarray}\label{letz}
&\fl\Re\left[C^*(k_0){\rm e}^{-ik_0x}{\rm e}^{i\hbar k_0^2/(2m)}\left\langle\Psi_f(x,t)\right\rangle\right]=-\frac{\alpha^2m^2}{2\hbar^4k_0^2}\Re\left\{C^*(k_0)A^*(k_0)B(k_0){\rm e}^{-2ik_0L_1}
\left\{C(k_0)\right.\right.\nonumber\\&\fl\left.\left.\times \left[A(k_0)B^*(k_0)\left({\rm e}^{2ik_0L_2-4\sigma^2k_0^2/\theta}+{\rm e}^{2ik_0L_1}\right)+A^*(k_0)B(k_0)
\left({\rm e}^{-2ik_0L_2}
+{\rm e}^{-2ik_0L_1-4\sigma^2k_0^2/\theta}\right)\right]\right.\right.\nonumber\\&\fl\left.\left.+R(k_0){\rm e}^{-ik_0(L_2-L_1)}{\rm e}^{-2\sigma^2k_0^2/\theta}\left(1+{\rm Erf}(i\sigma k_0/\sqrt{\theta})
\right)\left[A(k_0)B^*(k_0){\rm e}^{2ik_0L_2}+A^*(k_0)B(k_0)\right.\right.\right.\nonumber\\&\fl\left.\left.\left.\times{\rm e}^{-2ik_0L_1}
+A^*(k_0)B(k_0){\rm e}^{-2ik_0L_2}+A(k_0)B^*(k_0){\rm e}^{2ik_0L_1}
\right]+4R(k_0){\rm e}^{-ik_0(L_2-L_1)}\Im\left[\frac{k_0}{\tilde{k}_1^2-k_0^2}\right.\right.\right.\nonumber\\&\fl\left.\left.\left.\times{\rm e}^{-\sigma^2(\tilde{k}_1^2+k_0^2)/\theta}
\left(1+{\rm Erf}(i\sigma\tilde{k}_1/\sqrt{\theta})\right)\left(B^*(\tilde{k}_1)\frac{A_o(\tilde{k}_1)}
{\mathcal{D}_u'(\tilde{k}_1)}
{\rm e}^{2i\tilde{k}_1L_2}+A^*(\tilde{k}_1)\frac{B_o(\tilde{k}_1)}
{\mathcal{D}_u'(\tilde{k}_1)}{\rm e}^{-2i\tilde{k}_1L_1}\right)\right.\right.\right.\nonumber\\&\fl\left.\left.\left.+
\left(A^*(\tilde{k}_1)\frac{B_o(\tilde{k}_1)}{\mathcal{D}_u'(\tilde{k}_1)}{\rm e}^{-2i\tilde{k}_1L_2}+B^*(\tilde{k}_1)\frac{A_o(\tilde{k}_1)}{\mathcal{D}_u'(\tilde{k}_1)}{\rm e}^{2i\tilde{k}_1L_1}
\right)\right]\right\}\right\}
\end{eqnarray}
Again a complementary contribution is obtained by exchanging $A(k)$ and $B(k)$ and complex conjugating the $\sigma$-independent exponentials.

\section{Conclusions}
We studied the impact of noise on the propagation of waves in scattering systems. We considered as free parameters the wavenumber $k$, the strength of the perturbation $\alpha$ and the memory time of the noise determined by $\theta/\sigma^2$ and calculated the effect of the noise  on the current density through the system. Here, for modeling noise we add a potential $V(x,t)=\Theta(t)\delta(x-\gamma(t))$ and perturbatively calculate corrections to the current with respect to $\alpha$. The dynamics of $\gamma(t)$ is described by an Ornstein Uhlenbeck process.

We start by considering a free wave that is perturbed by a delta potential located at a time-dependent position.
In this case we obtain an exponential decay of the current correction induced by the perturbation with $\sigma^2/\theta$: Starting out with $\sigma^2=0$, i.e.\ a static setup, we get the current correction induced by a single static delta potential. With increasing $\sigma^2$, i.e.\ decreasing the memory time of the noise, this correction is reduced exponentially by the factor $\exp\left(-\sigma^2k_0^2/\theta\right)$. 
Afterwards we add the same noise to a quantum graph. Similar to \cite{Derev}, where the effect of nonlinearities on sharp resonances is studied, we find that the resonances are retained also under the influence of the perturbation and the shape and the position of the resonances depends in a complex way on the parameters $\alpha$, $k$ and $\sigma^2$ as described in the text. As in \cite{Derev} the effect of the perturbation is getting stronger with decreasing width of the unperturbed resonance.

One major drawback of our analysis is its perturbative nature. This restricts especially for very sharp resonances the regime of allowed values for $\alpha$ quite strongly. The regime of large values of $\alpha$ where especially strong deformations of the resonances are to be expected cannot be assessed. A nonperturbative method to analyze the impact of noise on resonances would thus be highly desirable.
Other possible extensions are  to consider the impact of noise on other properties of quantum graphs as on the spectral features of closed systems. Furthermore an extension of this calculation to higher dimensional systems would be desirable. In this context a comparison with the effect of noise found within other models \cite{Petitjean}, that considered the impact of the noise on the transmission expressed within a semiclassical framework, would be interesting.

 \section*{Acknowledgments}
\noindent  It is a pleasure to thank Rami Band, Stanislav Derevyanko and Michael Aizenman for instructive discussions and critical comments. This work was supported by
the Minerva Center for non-linear Physics, the Einstein (Minerva)
Center at the Weizmann Institute and the Wales Institute of
Mathematical and Computational Sciences (WIMCS) and The Israel Science Foundation  ISF - 861/11 (F.I.R.S.T.) grant 'Nonlinear waves and lasing on random networks'.
 D.W. acknowledges support by a Minerva fellowship.


\section*{References:}
\begin{thebibliography}{10}
\bibitem{KS} T.\ Kottos and U.\ Smilansky,  Phys.\ Rev.\ Lett.\ {\bf 79}, 4794- 4797, (1997); Annals of Physics {\bf 274}, 76-124 (1999).

\bibitem{Berkolaiko} G.\ Berkolaiko, H.\ Schanz, R.\ S.\ Whitney, Phys.\ Rev.\ Lett.\ {\bf 88}, 104101 (2002).

\bibitem{AltGnutz} A.\ Altland and S.\ Gnutzmann, Phys.\ Rev.\ Lett.\ {\bf93}, 194101 (2004).

\bibitem{KS-scat} T.\ Kottos and U.\ Smilansky, Phys.\ Rev.\ Lett.\ {\bf 85}
968, (2000); J. Phys. A. {\bf 36} 3501-3524 (2003).

\bibitem{Weidenmueller} Z. Pluhar and H.A. Weidenm\"uller, Phys.\ Rev.\ Lett.\ {\bf110}, 034101 (2013).

\bibitem{Bohigas} O.\ Bohigas, M.\ J.\ Giannoni, C.\ Schmit, Phys.\ Rev.\ Lett.\ {\bf 52}, 1 (1984).

\bibitem{Haake}  F. Haake, Quantum Signatures of Chaos,(3rd Edition)
Springer Heidelberg Dordrecht London New York, DOI 10.1007/978-3-642-05428-0 (2010).

\bibitem {Gnutzmann} S.\ Gnutzmann and U. Smilansky, Adv.\ in Phys.\ {\bf55}, 527 (2006).

\bibitem {BerkoKuch} G. Berkolaiko and P. Kuchment, Introduction to Quantum Graphs, Mathematical Surveys and Monographs 186; 270 pp; hardcover. AMS (2013).

\bibitem {BlumSmi}R.\ Bl\"umel and U.\ Smilansky,  Phys.\ Rev.\ Lett.\ {\bf 64}, 241 (1990).

\bibitem{Warsaw1} O.\ Hul, S.\ Bauch, P.\ Pakonski, N.\ Savytskyy, K.\ Zyczkowski, and L.\ Sirko, Phys.\ Rev.\ E {\bf69}, 056205 (2004).

\bibitem{Warsaw2} M.\ Lawniczak, S.\ Bauch, A.\ Sawicki,  M.\ Kus and L.\ Sirko, Acta Physica Polonica A {\bf 124}, 1078-1081,  (2013).

\bibitem{Marbourg} S.\ Gehler, U.\ Kuhl, H.-J.\ Stöckmann, (Private communication) (2014).

\bibitem{Darmstadt} B. Dietz (Private communication) (2014).

\bibitem{Nir} N. Davidson (Private communication) (2014).

\bibitem{Schan} S.\ Gnutzmann, H.\ Schanz, U.\ Smilansky, Phys.\ Rev.\ Lett.\ {\bf110}, 094101 (2013).

\bibitem{Waltner} D.\ Waltner, U.\ Smilansky, Acta Physica Polonica A {\bf 124},1087, (2013).

\bibitem {Uhlenbbeck} G.\ E.\ Uhlenbeck and L.\ S.\ Ornstein, Phys. Rev. {\bf 36} 823, (1930).

\bibitem{Wang+Uhl} M.\ C.\ Wang, G.\ E.\ Uhlenbeck, Revs. Mod. Phys. {\bf 17} 323, (1945).

\bibitem{Risken} H.\ Risken, {\it The Fokker-Planck Equation: Methods Of Solutions And Applications}, Springer, (1996); C.\ W.\ Gardiner, {\it Handbook of Stochastic Methods}, Springer, (2004).

\bibitem{Alder}
A.\ Winther, K.\ Alder, {\it Electromagnetic excitation: theory of Coulomb excitation with heavy ions}, Elsevier.


\bibitem{Leschke}
B.\ Bodmann, H.\ Leschke, S. Warzel, Path integrals: Dubna '96, eds. V. S. Yarunin and M. A. Smondyrev, pp. 95-106.


\bibitem{Band}
R.\ Band, G.\ Berkolaiko, U.\ Smilansky, Ann.\ Henri Poincar\'e {\bf 13}, 145 (2012).
\bibitem{Scatt}
M.\ L.\ Goldberger, K.\ M.\ Watson, {\it{Collision Theory}}, Wiley (1964); R.\ G.\ Newton, {\it{Scattering Theory of Waves and Particles}}, McGraw-Hill, New York (1966); 
C.\ J.\ Jochain, {\it{Quantum Collision Theory}}, North Holland (1983).

\bibitem{Derev}
S.\ Gnutzmann, U.\ Smilansky, S.\ Derevyanko, Phys.\ Rev.\ A {\bf 83}, 033831 (2011).

\bibitem{Petitjean}
C.\ Petitjean, P.\ Jacquod, R.\ Whitney, JETP Letters {\bf 86}, 647 (2008); A.\ Altland, P.\ W.\ Brouwer, C.\ Tian, Phys.\ Rev.\ Lett.\ {\bf99}, 036804 (2007);
F.\ M.\ Cucchietti, C.\ H.\ Lewenkopf, H.\ M.\ Pastawski, Phys.\ Rev.\ E {\bf74}, 026207 (2006).
\end{thebibliography}
\end{document}